\documentclass[12pt]{article}
\usepackage{amsmath}
\input{epsf}
\def\lra{\leftrightarrow}
\def\Ph{{\hat P}}
\def\zb{{\bar z}}

\newcommand\as{\alpha_{\mathrm{S}}}

\def\r{\sqrt{(1-z)^2-4z\, q_T^2/Q^2}}
\def\Li{\mathop{\hbox{\rm Li}}\nolimits}
\newcommand\f[2]{\frac{#1}{#2}}
\def\ep{\epsilon}
\def\qt{q_T}
\def\ms{${\overline {\rm MS}}$}
\def\beeq{\begin{eqnarray}}
\def\eeeq{\end{eqnarray}}
\def\cm{{\cal M}}
\def\oas{${\cal O}(\as)$}
\def\oass{${\cal O}(\as^2)$}
\def\qb{{\bar q}}
\def\cb{{\bar c}}

\def\to{\rightarrow}
\def\res{{\rm res.}}

\def\nn{\nonumber}

\def\pen{\frac{1}{(N+1)(N+2)} }
\def\funpn{{\cal F}_{c\cb}(N,\epsilon)}
\def\funqn{{\cal F}_{q{\bar q}}(N,\ep)}
\def\fungn{{\cal F}_{gg}(N,\ep)}
\def\funpq{{\cal F}_{q{\bar q}}(0,\epsilon)}
\def\funpg{{\cal F}_{gg}(1,\epsilon)}
\def\peng{\frac{1}{(N+2)} }
\def\K{{\cal K}}

\begin{document}
\begin{titlepage}
\renewcommand{\thefootnote}{\fnsymbol{footnote}}
\begin{flushright}
    hep-ph/0108273
     \end{flushright}
\par \vspace{10mm}
\begin{center}
{\huge
The structure of large logarithmic corrections \\[0.8ex]
at small transverse momentum  \\[1.5ex]
in hadronic collisions}
\footnote{This work was supported in part 
by the EU Fourth Framework Programme ``Training and Mobility of Researchers'', 
Network ``Quantum Chromodynamics and the Deep Structure of
Elementary Particles'', contract FMRX--CT98--0194 (DG 12 -- MIHT).}

\end{center}
\par \vspace{2mm}
\begin{center}
{\bf Daniel de Florian$\footnote{Partially supported by Fundaci\'on Antorchas and CONICET}
{}^{\,a,d}$}
\hskip .2cm
and
\hskip .2cm
{\bf Massimiliano Grazzini$\footnote{Partially supported by the Swiss National Foundation}
{}^{\,b,c,d}$}\\

\vspace{5mm}

${}^{a}\,$Departamento de F\'\i sica, FCEYN, Universidad de Buenos Aires,
(1428) Pabell\'on 1 Ciudad Universitaria, Capital Federal, Argentina

${}^{b}\,$Dipartimento di Fisica, Universit\`a di Firenze, I-50125 Florence, Italy 

${}^{c}\,$INFN, Sezione di Firenze, I-50125 Florence, Italy 

 ${}^{d}\,$Institute for Theoretical Physics, ETH-H\"onggerberg, 
 CH 8093 Z\"urich, Switzerland

\end{center}

\par \vspace{2mm}
\begin{center} {\large \bf Abstract} \end{center}
\begin{quote}
\pretolerance 10000

We consider the region of small transverse
momenta in the production of high-mass systems in hadronic collisions.
By using the current knowledge on the
infrared behaviour of tree-level and one-loop QCD amplitudes at \oass, we
analytically compute the general form of the logarithmically-enhanced contributions
up to next-to-next-to-leading logarithmic accuracy.
By comparing the results with $q_T$-resummation formulae
 we extract the coefficients that control the resummation of the large logarithmic contributions for both quark and gluon channels.
Our results show that within the conventional resummation formalism the Sudakov form factor is actually process-dependent.

\end{quote}

\vspace*{\fill}
\begin{flushleft}
     hep-ph/0108273 \\August 2001 

\end{flushleft}
\end{titlepage}

\renewcommand{\thefootnote}{\arabic{footnote}}
\setcounter{footnote}{0}

\section{Introduction}
\label{sec:intro}

The transverse-momentum distribution of systems with high invariant mass
produced in high-energy hadron collisions is important for QCD studies
and for physics studies beyond the Standard Model
(see, e.g., Refs.~\cite{Catani:2000jh}--\cite{Baur:2000xd}).

We consider the inclusive hard-scattering process
\begin{equation}
\label{class}
h_1(p_1) + h_2(p_2) \to F(Q^2,\qt^2;\phi) + X \;\;,
\end{equation}
where the final-state system $F$ is produced by the collision of the
two hadrons $h_1$ and $h_2$ with momenta $p_1$ and  $p_2$, respectively.
The final state $F$ is a generic system of non-strongly interacting particles,
such as {\em one} or {\em more} vector bosons $(\gamma^*, W, Z, \dots)$, 
Higgs particles ($H$) and so forth. 
We denote by $\sqrt s$ the center-of-mass energy of the colliding hadrons
$(s= (p_1+p_2)^2 \simeq 2p_1p_2)$, and by $Q^2$ and $q_T^2$ the invariant mass
and total transverse momentum of the system $F$, respectively.
The additional variable $\phi$ in (\ref{class}) denotes the possible dependence on the kinematics
of the final state particles in $F$ (such as rapidities,
 individual transverse momenta and so forth).

We assume that at the parton level the system $F$ is produced with
vanishing $\qt$ (i.e. with no accompanying final-state radiation)
in the leading-order (LO) approximation.
Since $F$ is colourless, the LO partonic subprocess is either
$q_f \bar{q}_{f'}$ {\em annihilation}, as in the case of $\gamma^*, W$ and $Z$ 
production, or $gg$ {\em fusion}, as in the case of the production of a Higgs
boson $H$.

When the transverse momentum of the produced system $\qt^2$ is of the order of its invariant mass $Q^2$ the fixed order calculation is reliable\footnote{It is assumed that all other dimensionful invariants are of the same
  order $Q^2$.}.
In the region $\qt^2\ll Q^2$ large logarithmic corrections of the form 
$\as^n/\qt^2 \log^{2n-1} Q^2/\qt^2$ appear,
 which spoil the convergence of fixed-order perturbative calculations. The
logarithmically-enhanced terms have to be evaluated at higher
perturbative orders, and possibly resummed to all orders in the QCD coupling
constant $\as$. The all-order resummation formalism was developed in the eighties
\cite{Dokshitzer:1980hw}--\cite{ds}.
The structure of the resummed distribution is given in terms of a 
transverse-momentum form factor and of process-dependent contributions.

The coefficients that control the resummation of the large
logarithmic contributions for a given process in (\ref{class}) can be computed
at a given order 
if an analytic calculation at large $q_T$ at the same order exists.
At first order in $\as$ the structure of the large logarithmic contributions
is known to be universal and depends only on the channel in which the system is produced in the LO approximation. 
At second relative order in $\as$, only a few analytical calculations are available, like the pioneering one for
lepton-pair
Drell--Yan production, performed by Ellis, Martinelli and Petronzio in Ref.~\cite{Ellis:1983hk}.
Using the results of Ref.~\cite{Ellis:1983hk} Davies and Stirling \cite{ds} (see also \cite{Davies:1985sp}) 
were able to obtain the complete structure of the \oass\
logarithmic corrections for that process.

The analysis performed by Davies and Stirling is by far non trivial because it requires
the  integration of the analytic $\qt$ distribution in the small $\qt$ limit.
Moreover the calculation cannot tell anything about the dependence of these coefficients
on the particular process in (\ref{class}) and
should in principle
be repeated for each process.

In this paper we address this problem with a completely independent
and general method. Our basic observation is that the
large logarithmic corrections are of infrared (soft and collinear) nature, and thus
their form can be predicted once and for all
in a general (process independent) manner.

The structure of the logarithmically-enhanced contributions at ${\cal O}(\as^n)$ is controlled
by the infrared limit of the relevant QCD amplitudes at the same order.
The infrared behaviour of QCD amplitudes at ${\cal O}(\as)$ is known since
long time \cite{Bassetto:1983ik}.
Recently, soft and collinear singularities arising in tree-level \cite{Campbell:1998hg,Catani:2000ss} and
one-loop \cite{dixon,Bern:1998sc,Kosower:1999rx,Catani:2000pi} QCD amplitudes at \oass\
have been extensively studied and the corresponding kernels
have been computed \cite{Campbell:1998hg}-\cite{DelDuca:2000ha}.
By using this knowledge, and exploiting the
relatively simple
kinematics
of the process (\ref{class}), we will construct
general approximations of the relevant
QCD matrix elements that are able to control all singular regions corresponding to $\qt\to 0$ avoiding double counting.
By using these approximations we will compute the general structure of
the ${\cal O}(\as^2)$-logarithmically-enhanced contributions
both for ${\bar q}q$- and for $gg$-initiated processes.

The results
provide an important check of the validity of the resummation formalism
and allow to extract the general form of the resummation coefficients.
In particular in the quark channel we can confirm the
results of Ref.~\cite{ds} in the case of Drell--Yan and in the gluon channel
we can give the coefficients in the important case
of Higgs boson production through gluon-gluon fusion.

The universality of our method relies on the fact that the infrared factorization formulae
we use
depend only
on the channel ($q{\bar q}$ or $gg$) in which the system $F$ is
produced at LO and not on the details of $F$.

Our main results were anticipated in a short letter \cite{deFlorian:2000pr}.
 This paper is organized as follows.
In section 2 we review the framework of the resummation formalism and present the
strategy for the calculation. In section 3 we perform the calculation explicitly
for the ${\cal O}(\as)$ corrections
and extract the first order coefficients. Section 4 and 5 are devoted to
the calculation at ${\cal O}(\as^2)$ for the quark and the gluon channel and constitute the main part of this work.
Finally in section 6 we present our final results and discussion. 

\section{Resummation formula}
\label{sec:resummation}

The transverse momentum distribution for the process in Eq.~(\ref{class}) can be written as:
\begin{equation}
\label{Fdec}
\frac{d\sigma_{F}}{dQ^2 \;d\qt^2\;d\phi} = 
\left[ \frac{d\sigma_{F}}{dQ^2 \;d\qt^2\;d\phi} \right]_{\res}
+ \left[ \frac{d\sigma_{F}}{dQ^2 \;d\qt^2\;d\phi} \right]_{{\rm fin.}} \;\;.
\end{equation}
Both terms on the right-hand side are obtained as convolutions of
partonic cross sections and the parton distributions $f_{a/h}(x,Q^2)$ 
($a=q_f, {\bar q}_f, g$ is the parton label) of the
colliding hadrons\footnote{Throughout the paper we always use
parton densities as defined in the \ms\ factorization scheme and $\as(q^2)$ is
the QCD running coupling in the \ms\ renormalization scheme.}.

The partonic cross section that enters in the resummed part (the first term
on the right-hand side) contains all the logarithmically-enhanced
contributions $\as^n/\qt^2 \log^m Q^2/\qt^2$. Thus, this part has to be evaluated by 
resumming the logarithmic terms to all orders in perturbation theory. On the
contrary, the partonic cross section in the second term on the right-hand side
is finite (or at least integrable) order-by-order in perturbation theory when $\qt \to 0$. It can thus
be computed by truncating the perturbative expansion at a given fixed order in
$\as$. 

Since in the following we are interested in the small-$\qt$ limit we will
be concerned only with
the first term in Eq.~(\ref{Fdec}).
The resummed component is\footnote{This expression can be generalized
to include the dependence on the renormalization and factorization scales 
$\mu_R$ and $\mu_F$, respectively (see e.g. Ref.~\cite{Catani:2001vq}).}
\begin{align}
\label{resgen}
\left[ \frac{Q^2\, d\sigma_{F}}{dQ^2 \;d\qt^2\,d \phi} \right]_{\res} =& \sum_{a,b}
\int_0^1 dx_1 \int_0^1 dx_2 \int_0^\infty db \; \frac{b}{2}\; J_0(b \qt) 
\;f_{a/h_1}(x_1,b_0^2/b^2) \; f_{b/h_2}(x_2,b_0^2/b^2) \nn \\
\cdot\,& s\, W_{ab}^{F}(x_1 x_2 s; Q, b, \phi) \;\;.
\end{align}
The Bessel function $J_0(b \qt)$ and the coefficient $b_0=2e^{-\gamma_E}$
($\gamma_E=0.5772\dots$ is the Euler number) have a kinematical origin.
To correctly take into account the kinematics constraint of transverse-momentum
conservation, the resummation procedure has to be carried out in the 
impact-parameter $b$-space.
The resummed
coefficient $W_{ab}^{F}$
is
\begin{align}
\label{nonunw}
W_{ab}^{F}(s; Q, b, \phi) &= \sum_c \int_0^1 dz_1 \int_0^1 dz_2 
\; C_{ca}^{F}(\as(b_0^2/b^2), z_1) \; C_{{\bar c}b}^{F}(\as(b_0^2/b^2), z_2)
\; \delta(Q^2 - z_1 z_2 s) \nn \\
&\cdot\, \f{d\sigma_{c{\bar c}}^{(LO) \,F}}{d\phi} \;S_c^{F}(Q,b) \;\;.
\end{align}
where $d\sigma_{c\bar{c}}^{(LO)}/d\phi$ corresponds to the
leading order
cross section for the production of the large invariant mass system $F$ in the $c{\bar c}$ channel, with $c$  representing either a quark $q$ or a gluon $g$.
The resummation of the large logarithmic corrections is achieved by 
exponentiation, that is by showing that the  
Sudakov form factor can be expressed as
\begin{equation}
\label{sudakov}
S_c(Q,b)=\exp \left\{ -\int_{b_0^2/b^2}^{Q^2} \frac{dq^2}{q^2} 
\left[ A_c(\as(q^2)) \;\log \frac{Q^2}{q^2} + B_c(\as(q^2)) \right] \right\}\, .
\end{equation}
 
The functions $A_c(\as)$, $B_c(\as)$, as well as the
coefficient functions $C_{ab}(\as,z)$ in Eqs.~(\ref{nonunw},\ref{sudakov})
are free of large logarithmic corrections and  have perturbative
expansions in $\as$ as
\begin{eqnarray}
\label{aexp}
A_c(\as) &=& \sum_{n=1}^\infty \left( \frac{\as}{2\pi} \right)^n A_c^{(n)} \;\;, \\
\label{bexp}
B_c(\as) &= &\sum_{n=1}^\infty \left( \frac{\as}{2\pi} \right)^n B_c^{(n)}
\;\;, \\
\label{cexp}
C_{ab}(\as,z) &=& \delta_{ab} \,\delta(1-z) + 
\sum_{n=1}^\infty \left( \frac{\as}{2\pi} \right)^n C_{ab}^{(n)}(z) \;\;.
\end{eqnarray}

The coefficients of the perturbative expansions $A_c^{(n)}$, $B_c^{(n)}$ and 
$C_{ab}^{(n)}(z)$ are the key of the resummation procedure since their knowledge allows to perform the resummation to a given {\it Logarithmic} order: 
 $A^{(1)}$ leads to the resummation of leading logarithmic (LL) contributions, $\{ A^{(2)}, B^{(1)}, C^{(1)} \}$ give the next-to-leading logarithmic (NLL)
terms, $\{ A^{(3)}, B^{(2)}, C^{(2)}\}$ give the 
next-to-next-to-leading logarithmic (NNLL) terms,
and so forth\footnote{In a different classification
the coefficient $C^{(1)}$ enters only at NNLL \cite{Frixione:1999dw}.}.
The coefficient functions $C_{ab}^{(n)}(z)$ depend on the process, as it has been confirmed by  calculations of $C_{ab}^{(1)}(z)$ for several processes.
The Sudakov form factor $S_c(Q,b)$ that enters Eq.~(\ref{nonunw}) is often 
{\it supposed} to be universal. However, as we will show, this is not 
the case, and anticipating our results we label all process-dependent 
coefficients by the upper index $F$.
 The 
coefficients $A^{(1)}$, $B^{(1)}$, $A^{(2)}$  are
universal and are
known both for
the quark \cite{Kodaira:1982nh} and for the gluon \cite{Catani:1988vd} 
form factors
\begin{align}
&A_q^{(1)} =2 C_F \;\;, \quad \quad \;\;A_g^{(1)} = 2 C_A \;, \nn\\
\label{ffcoef}
&B_q^{(1)} = - 3 C_F \;\;, \quad 
B_g^{(1)} = - 2\beta_0\;, \\
&A_q^{(2)} = 2 C_F K \;\;, \quad A_g^{(2)} = 2 C_A K \nn\;,
\end{align}
where
\begin{equation}
\beta_0=\f{11}{6}C_A-\f{2}{3}\,n_f T_R
\end{equation}
and
\begin{equation}
\label{K}
K = \left( \frac{67}{18} - \frac{\pi^2}{6} \right) C_A - \f{10}{9}\, n_f T_R \;\;.
\end{equation}
The NNLL coefficient $B^{(2)}$
was computed by Davies and Stirling \cite{ds} for the case of Drell--Yan (DY):
\begin{equation}
\label{b2dy}
B_q^{(2)DY}=C_F^2\left(\pi^2-\f{3}{4}-12\zeta(3)\right)+C_F\,C_A\left(\f{11}{9}\pi^2-\f{193}{12}+6\zeta(3)\right)+C_F\,n_f\,T_R\left(\f{17}{3}-\f{4}{9}\pi^2\right)\, ,
\end{equation}
where $\zeta(n)$ is the Riemann $\zeta$-function $(\zeta(3)=1.202\dots)$.
It is also worth noticing that, even though there is no analytical result available for it, the coefficient $A_{q,g}^{(3)}$ has been extracted numerically with a very good precision in Ref.~\cite{vogtA3}. 

As anticipated in the introduction, a direct way to obtain the coefficients
in Eqs.~(\ref{aexp}-\ref{bexp}) at a given order involves the  computation of
 the differential cross section ${d\sigma}/{dq_T^2 dQ^2 d\phi}$
 at small $q_T$ at the same order.
A comparison with the power expansion in
$\as$ of the resummed result in Eq.~(\ref{resgen}) allows
to extract the coefficients that
control the resummation of the large logarithmic terms. 
However, it has been shown by Davies and Stirling that is it more convenient
 to take $z=Q^2/s$ moments\footnote{Here we follow Ref.~\cite{ds} in the unconventional definition
of the moments:
$f(N)=\int_0^1 dz z^N f(z)$.}
of the  differential cross section defining
 the dimensionless quantity
\begin{equation}
\label{sigman}
\Sigma(N)=\int_0^{1-2 q_T/Q} dz\, z^N \f{Q^2 q_T^2}{d\sigma_0/d\phi}\f{d\sigma}{dq_T^2
  dQ^2d\phi}\, .
\end{equation}
Notice that in the definition of $\Sigma$ 
the cross section has been normalized
with respect to the lowest order partonic contribution ${d\sigma_0/d\phi}$ and  multiplied by $q_T^2$ to cancel its $1/q_T^2$ singular behaviour in the limit  $q_T \to 0$.
The upper limit of integration 
$z=1-2 q_T/Q (\sqrt{1+q_T^2/Q^2}-q_T/Q)\sim 1-2 q_T/Q $ 
has been approximated to a first order expansion in $q_T/Q$ and corresponds
 to the kinematics for  the emission of soft particles (i.e, when the center of mass
 energy $s$ is just enough to produce the system with invariant mass $Q$ and
 transverse momentum $q_T$).
Working with moments
allows to avoid complicated convolution integrals implicit in
(\ref{resgen}) and
makes possible to
factorize the parton densities from the partonic contribution
to the cross section.
In this way, the corresponding expression from the resummed formula 
(\ref{resgen}) reads
\begin{equation}
\Sigma(N)= \sum_{i,j} f_{i/h_1}(N,\mu^2_F) f_{j/h_2}(N,\mu^2_F)\,
\Sigma_{ij}(N)
\end{equation}
where 
\begin{align}
\label{sigman2}
&\Sigma_{ij}(N)= \sum_{a,b}
 \int_0^\infty b\,db \f{q_T^2}{2}\, J_0(b q_T)  \,
{C}^{{F}}_{ca}\left(N,\as\left({b_0^2}/{b^2} \right)\right)
{C}^{{F}}_{\overline{c}b}\left(N,\as\left({b_0^2}/{b^2} \right)\right)
\nn\\
&\cdot \exp \Bigg\{\! -\int_{b_0^2/b^2}^{Q^2} \frac{dq^2}{q^2} 
\left[ {A}_c(\as(q^2)) \;\log \frac{Q^2}{q^2} + {B}_c^{{F}}(\as(q^2
)) \right]
- \!\int_{{b_0^2}/{b^2}}^{\mu^2_F} \f{dq^2}{q^2} 
   \, \left( \gamma_{ai}+\gamma_{bj}\right) (N,\as(q^2)) \Bigg\}\nn\\
\end{align}
and an ordered exponential is understood.
Notice that the appearance of an extra term involving the
anomalous dimensions $\gamma_{ab}$
in the exponential in (\ref{sigman2})
is due to the
 evolution of the parton densities from the  original scale $b_0^2/b^2$ in
 (\ref{resgen}) to the arbitrary factorization scale $\mu_F$ at which they 
are now evaluated.

In order to extract the resummation coefficients, we can directly
 study  the partonic contribution $\Sigma_{ij}$. Furthermore, since we
want to perform a calculation of $\Sigma_{ij}$ to \oass \, and our main interest is the second order coefficient $B^{(2)}$, it is clear that only the diagonal contribution $\Sigma_{c{\bar c}}$ can give the desired information.
Each possible ``flavour changing'' contribution in Eq. (\ref{sigman2})
would add at least one extra power of $\as$ in the perturbative expansion.
`Non-diagonal' contributions to $\Sigma_{ij}$, which can be evaluated in a simpler way, might be used to check the structure  and consistency of 
the resummation framework at a given perturbative order but do not provide any additional information on the coefficients.

In order to have transverse momentum $q_T\ne 0$ at least one gluon has to be emitted and, therefore, the perturbative expansion of $\Sigma_{c\overline{c}}$
begins at ${\cal O}(\as)$
\begin{equation}
\label{sigmaexp}
\Sigma_{c\overline{c}}(N)=\f{\as}{2\pi}\;\Sigma^{(1)}_{c\overline{c}}(N)+\left(\f{\as}{2\pi}\right)^2\Sigma^{(2)}_{c\overline{c}}(N)+\cdots \,.
\end{equation}
From the expansion of the resummed formula (\ref{sigman2}) it is possible to
 obtain the expression for the first two coefficients in (\ref{sigmaexp}) as\footnote{For the sake of simplicity in the presentation, and unless otherwise stated,  we fix the 
factorization and renormalization scales to $\mu_F^2=\mu_R^2=Q^2$.}
\begin{eqnarray}
\label{sigma1}
\Sigma_{c{\bar c}}^{(1)}(N)={A}_c^{(1)} \log\f{Q^2}{q_T^2} +{B}_c^{(1)} +2 \gamma_{cc}^{(1)}(N) 
\end{eqnarray}
and 
\begin{align}
\label{sigma2}
&{\Sigma^{(2)}_{c\bar{c}}}{(N)}
= \log^3\f{Q^2}{q_T^2} \left[-\f{1}{2} \left( {A}_c^{(1)}\right)^2 \right]
 \nn\\&
+ \log^2\f{Q^2}{q_T^2} \left[-\f{3}{2} \left( {B}_c^{(1)} +2 \gamma_{cc}^{(1)}(N)
  \right) {A}_c^{(1)} + \beta_0 {A}_c^{(1)} \right] \nn\\
&+ \log\f{Q^2}{q_T^2} \left[ {A}_c^{(2)} +
 \beta_0 \left({B}_c^{(1)} + 2  \gamma_{cc}^{(1)}(N)  \right) - 
\left(  {B}_c^{(1)} +2 \gamma_{cc}^{(1)}(N) \right)^2
  \right. \nn\\ &   \hspace{1.2cm} \left. 
+ 2 {A}_c^{(1)} {C}_{cc}^{(1){F}}(N)  
-2 \sum_{j\ne c} \gamma_{cj}^{(1)}(N) \gamma_{jc}^{(1)}(N)
\right] \nn\\
&+ {B}_c^{(2){F}} + 2 \gamma_{cc}^{(2)}(N)
 +
  2 \left( {B}_c^{(1)} + 2 \gamma_{cc}^{(1)}(N)\right)
 {C}_{cc}^{(1){F}}(N)
  +2 \zeta(3) ({A}_c^{(1)})^2
\nn \\ & 
- {2 \beta_0} {C}_{cc}^{(1){F}}(N) 
+ 2 \sum_{j\ne c} \left[ {C}_{cj}^{(1){F}}(N) \gamma_{jc}^{(1)}(N)
%
%
 \right] \, .
\end{align}

The computation of $\Sigma_{c{\bar c}}^{(1)}(N)$ can provide information
on the first order coefficients ${A}_c^{(1)}$ (the logarithmic term
in (\ref{sigma1})) and ${B}_c^{(1)}$ (the constant term in (\ref{sigma1}))
as well as on the
one-loop anomalous dimensions $\gamma_{cc}^{(1)}(N)$ ( the $N-$dependent term
in (\ref{sigma1})\footnote{Notice that all moments but one 
 can  actually be extracted. The remaining one can be obtained by imposing  quark number and momentum conservation rules.}.).
In the same way, the coefficients ${A}_c^{(2)}$ and ${B}_c^{(2){F}}$ can be extracted from the second order result (\ref{sigma2}). At this order, also the coefficient functions  ${C}_{ij}^{(1){F}}(N)$ contribute to the logarithmic and constant terms and therefore should be known in order to be able to proceed with the extraction of ${A}_c^{(2)}$ and ${B}_c^{(2){F}}$ . Fortunately,  there is another related quantity which allows to obtain the coefficient functions ${C}_{ij}^{(1){F}}(N)$ from a first order calculation. 
 This is the $q_T$-integrated cross section
\begin{equation}
\label{inteqt}
\int_0^{p_T^2} \f{dq_T^2}{q_T^2} \Sigma_{i\overline{c}}\, .
\end{equation}
When $p_T^2\ll Q^2$ the perturbative expansion to ${\cal O}(\as)$ reads (neglecting again terms that vanish when $p_T\to 0$)
\begin{align}
\label{inqt}
\int_0^{p_T^2} \f{dq_T^2}{q_T^2} \Sigma_{c\overline{c}}&=
\f{\as}{2\pi} \left[ -\f{1}{2} {A}_c^{(1)} \log^2\f{Q^2}{p_T^2} -
  ({B}_c^{(1)} +2 \gamma_{cc}^{(1)}(N)) \log\f{Q^2}{p_T^2} + 2
 {C}_{cc}^{(1){F}}(N)
\right] \nn \\
\int_0^{p_T^2} \f{dq_T^2}{q_T^2} \Sigma_{i\overline{c}}&=
\f{\as}{2\pi} \left[  - \gamma_{ci}^{(1)}(N) \log\f{Q^2}{p_T^2} + 
 {C}_{ci}^{(1){F}}(N)\right]   ~~~~~~~~~~~~~~~~~~  i \ne c \, .
\end{align}
The integration over $q_T$ adds one power in the  logarithm, with the coefficient functions  ${C}_{ij}^{(1){F}}(N)$ appearing now in the constant term. It is worth noticing that at variance with the calculation of $\Sigma$ the configuration with $q_T=0$ now contributes to Eq.~(\ref{inteqt}).

In the quark channel $(c=q)$, for the sake of simplicity and in order to 
compare directly with the calculation performed in \cite{ds}, we will
concentrate on
the {\it non-singlet}
contribution
to the cross section defined by
\begin{equation}
\label{eq:nonsing}
\sigma^{\rm NS}= \sum_{ff'} \left( \sigma_{q_f \bar{q}_{f'}} - \sigma_{q_f q_{f'}}
\right)\, .
\end{equation}
The  second order expansion for $\Sigma_{q\overline{q}}^{\rm NS}(N)$
in terms of the resummation coefficients reads like the one in Eq.~(\ref{sigma2}) but without the `singlet' contributions involving $ \sum_{j\ne c}$ and with the corresponding
non-singlet
anomalous dimension.
In the following the label NS will be always understood in $\Sigma_{q{\bar q}}$.

\section{The calculation at \oas}
\label{sec:oas}

The calculation at ${\cal O}(\as)$ is not difficult and the results are
 rather well known. Nevertheless, we will give in this section the details on the computation as a way  to present the 
main ideas of the method developed to obtain the resummation coefficients. 

\begin{figure}[htb]
\begin{center}
\begin{tabular}{c}
\epsfxsize=6truecm
\epsffile{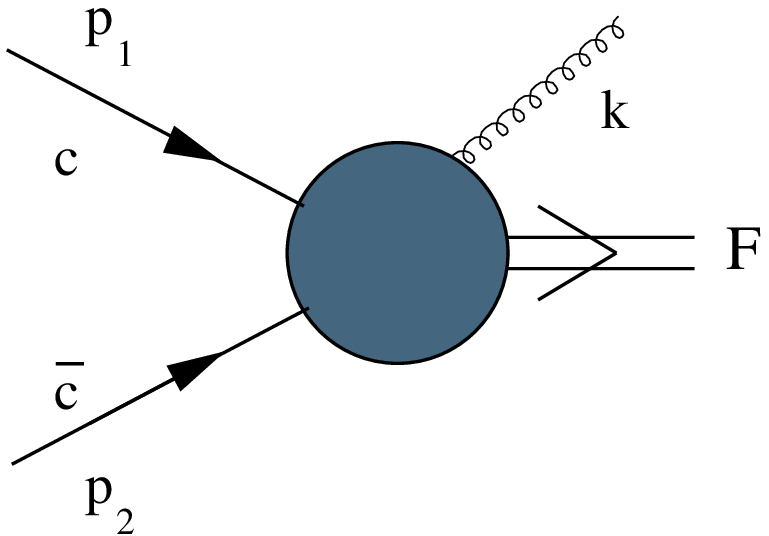}\\
\end{tabular}
\end{center}
\caption{\label{fig:1}{\em \oas\ contribution to the process (1) }}
\end{figure}

At this order only one extra gluon of momentum $k$ can be radiated and the
kinematics for the process $c\bar{c}\to g+F $ is (see Fig.~\ref{fig:1})
\begin{equation}
p_1+p_2\to k+q\, .
\end{equation}

We denote the corresponding matrix element by 
${\cal M}^{(0)}_{c\bar{c}\to
    g\, F}\left(p_1,p_2,k,\phi\right)$
and the usual invariants are defined as
\begin{equation}
s=(p_1+p_2)^2~~~~u=(p_2-k)^2~~~~t=(p_1-k)^2~~~~z=Q^2/s\, .
\end{equation}
The differential cross section can be written as
\begin{equation}
\label{dsigma1}
\f{d\sigma_{c\cb\to g\, F}}{dq_T^2 dQ^2d\phi}= \int \f{|{\cal M}^{(0)}_{c\bar{c}\to
    g\, F}\left(p_1,p_2,k,\phi\right)|^2}{8 s (2\pi)^2}\f{(4\pi)^\ep
  q_T^{-2\ep}}{\Gamma(1-\ep)}\,  \f{du}{u}\, \delta\left( \f{1}{u}
  (u-u_{max})(u-u_{min}) \right) \, ,
\end{equation}
where the two roots of the equation $(p_1+p_2-q)^2=0$ are given by
\begin{equation}
\label{umaxumin}
u_{min}=Q^2\;\f{z-1-\r}{2z}~~~~~~~~~u_{max}=Q^2\;\f{z-1+\r}{2z} \, .
\end{equation}
In order to regularize both ultraviolet and infrared divergences we 
work in the
conventional dimensional regularization scheme (CDR) with $4-2\ep$ space-time dimensions, considering two helicity states for 
massless quarks and $2-2\ep$ helicity states for gluons.
The lowest-order cross section (at $q_T=0$) needed to construct $\Sigma$ in 
Eq.~(\ref{sigman}) is given by
\begin{equation}
\label{eqps1}
\f{d\sigma_0}{d\phi}=\f{|{\cal
    M}_{c{\bar c}\to F}^{(0)}\left(p_1,p_2,\phi\right)|^2}{2s} \, ,
\end{equation}
in terms of the Born matrix element $|\cm_{c\cb\to F}^{(0)}\left(p_1,p_2,\phi\right)|^2$.
 
As has been stated, we  want to obtain $\Sigma^{(1)}_{c{\bar c}}$
by using our knowledge on the
behaviour of QCD matrix elements in the soft and collinear regions at
\oas .
The starting point is the observation that, when $\qt^2$ is small, the additional gluon is constrained to be either collinear to one of the incoming partons
or soft.
Thus there are three singular regions of ${\cal M}^{(0)}_{c\cb\to F}
\left(p_1,p_2,k,\phi\right)$ in the $q_T \to 0$ limit:
\begin{itemize}
\item first collinear region: $p_1k\to0$
\item second collinear region: $p_2k\to0$
\item soft region: $k\to 0$.
\end{itemize}
It is clear that, since $\qt^2$ is small but does not vanish, these regions do not produce any real singularity, i.e. poles in $\epsilon$, but are responsible for the appearance of the
logarithmically-enhanced contributions.
When $p_1k\to 0$ the matrix element squared factorizes as follows:
\begin{equation}
\label{col1}
|{\cal M}^{(0)}_{c\cb\to g\, F}\left(p_1,p_2,k,\phi\right)|^2\simeq \f{4\pi\as\mu^{2\ep}}{z_1p_1k}{\hat
  P}_{cc}(z_1,\ep) |{\cal M}^{(0)}_{c\cb\to F}\left(z_1\,p_1,p_2,\phi\right)|^2 \, ,
\end{equation}
where
\begin{align}
\label{pqq}
{\hat P}_{qq}(z,\ep)&=C_F\left[\f{1+z^2}{1-z}-\ep(1-z)\right]\\
\label{pgg}
\hat{P}_{gg}(z,\ep)&= 2 C_A \left[ \f{z}{1-z}+\f{1-z}{z}+z(1-z)\right] 
\end{align}
are the $\ep$-dependent real Altarelli--Parisi (AP) kernels in the CDR scheme.
In the left hand side of Eq.~(\ref{col1}) the matrix element squared
is obtained  replacing the two collinear partons $c$ and $g$ by
a parton $c$ with momentum $z_1 p_1$.

Notice that in the gluon channel there are additional spin-correlated contributions and Eq.~(\ref{col1}) is strictly valid only after azimuthal integration.
Since here and in the following we will always be interested in azimuthal integrated quantities, Eq.~(\ref{col1}) can be safely used also in the gluon channel.

In the limit $p_2k\to 0$ the singular behaviour is instead
\begin{equation}
\label{col2}
|{\cal M}^{(0)}_{c\cb\to g\,F}\left(p_1,p_2,k,\phi\right)|^2\simeq \f{4\pi\as\mu^{2\ep}}{z_2p_2k}{\hat
  P}_{cc}(z_2,\ep) |{\cal M}^{(0)}_{c\cb\to F}
\left(p_1,z_2\,p_2,\phi\right)|^2 \, .
\end{equation}
Let us now consider the limit in which the gluon becomes soft.
As it is well known soft-factorization formulae usually involve
colour correlations, that make colour and kinematics entangled.
In general  colour correlations relate each pair of hard momentum
partons in the Born matrix element.
In this case the hard momentum partons are only two and colour
conservation can be exploited to obtain:
\begin{equation}
\label{soft}
|{\cal M}^{(0)}_{c\cb\to g\,F}\left(p_1,p_2,k,\phi\right)|^2\simeq
4\pi\as\mu^{2\ep}\;C_c\, 4\, {\cal S}_{12}(k)|{\cal M}^{(0)}_{c\cb\to F}\left(p_1,p_2,\phi\right)|^2 \, ,
\end{equation}
where
\begin{equation}
\label{eik}
{\cal S}_{12}(k)=\f{p_1p_2}{2p_1k\, p_2k}
\end{equation}
is the usual eikonal factor and
we have defined
\begin{equation}
C_q=C_F
~~~~~~~~C_g=C_A\, .
\end{equation}
In Eq.~(\ref{soft}) colour correlations are absent and
factorization is exact. This feature will persist also at \oass.
 
In each of the singular regions discussed above, 
Eqs.~(\ref{col1}), (\ref{col2}) and (\ref{soft}) provide an approximation
of the exact matrix element that can be used to compute the cross section
in the small $\qt$ limit.
In principle it might be possible to split the phase space integration
in regions where only soft or collinear configurations can arise,
and use
in each region the corresponding
approximation.
Unfortunately, such method probes to be very difficult to be extended
to \oass, where the pattern of singular configurations
is much more complicated.
Thus our strategy is  to unify the factorization formulae in order to
obtain an approximation that it is valid in the full phase space.

As can be easily checked, if we identify the momentum fractions
$z_1$ and $z_2$
with $z$,
the collinear factorization formulae in Eqs.~(\ref{col1},\ref{col2}) contain the correct soft limit in Eq.~({\ref{soft}). Therefore, the unification of soft and collinear limits is rather simple: the usual collinear factorization formula already contains both. Strictly speaking, one can use the symmetry in the initial states in order to perform the
 integration in Eq.~(\ref{dsigma1}) only over half of the phase space (i.e. by taking for instance only $u=u_{max}$) and multiplying the result by two. In this way only one possible collinear configuration can occur and Eq.~(\ref{col1}) provides the needed approximation for the matrix element. 

At this order it is even possible to write down a general factorization formula for the three configurations that shows explicitly the $1/q_T^2$ singularity of the matrix element squared as
\begin{equation}
\label{qtfact}
|{\cal M}^{(0)}_{c\cb\to g\,F}\left(p_1,p_2,k,\phi\right)|^2\to
\f{4\pi\mu^{2\ep}\as}{q_T^2}\f{2(1-z)}{z}{\hat
  P}_{cc}(z,\ep)|{\cal M}^{(0)}_{c\cb\to F}\left(\phi\right)|^2 \, ,
\end{equation}
where we have used Lorentz invariance in order to write
$|{\cal M}_{c\cb\to F}^{(0)}(\phi)|^2$
only as a function of the final state kinematics.
We can now use this formula to compute the small $q_T$ behaviour of
$\Sigma_{c\cb}(N)$ in a completely process independent manner.
In fact the process dependence, given by the Born matrix element, is
completely factored out and cancels in $\Sigma$.
By replacing Eq.~(\ref{qtfact}) in Eq.~(\ref{dsigma1}) and using the definition of $\Sigma$
we obtain, keeping for future use its $\epsilon$ dependence:
\begin{align}
\label{sigma1p}
\Sigma^{(1)}_{c\cb}(N,\epsilon)&= \f{1}{\Gamma(1-\ep)}
\left(\f{4\pi \mu^2}{q_T^2}\right)^\ep\int_0^{1-2 q_T/Q} dz\,z^N \f{2(1-z)\hat
  {P}_{cc}(z,\epsilon) }{\r} \nn\\
&\equiv \f{1}{\Gamma(1-\ep)}\left(\f{4\pi \mu^2}{q_T^2}\right)^\ep 
C_c\,\funpn \, .
\end{align}
Explicitly, setting $\epsilon$ to 0, we have 
\begin{eqnarray}
\Sigma_{q{\bar q}}^{(1)}(N)=2C_F\,\log
\f{Q^2}{q_T^2}-3C_F+2\gamma_{qq}^{(1)}(N)\nn  \\
\Sigma_{gg}^{(1)}(N)=2C_A\log\f{Q^2}{q_T^2}-2\beta_0+2\gamma_{gg}^{(1)}(N) 
\, ,
\end{eqnarray}
for the quark and gluon channels.
Comparing to  Eq.~(\ref{sigma1}) we see that
$A^{(1)}_c=2 C_c$ is the coefficient of the leading $1/(1-z)$ singularity in the AP splitting functions
 whereas 
$B^{(1)}_c= -2 \gamma^{(1)}_c$ is given by the coefficient of the delta function in the regularized AP kernels 
\begin{eqnarray}
\gamma_q^{(1)}= \f{3}{2} C_F \,\,\,\,\,\,\,\,\,\, \gamma_g^{(1)}= \beta_0\, .
\end{eqnarray}

Finally, in order to obtain the coefficient $C_{ab}^{(1)}$, we have to evaluate
the integrals in Eq.~(\ref{inteqt}) and compare to the results
from Eqs.~(\ref{inqt}).
As far as the diagonal contribution is concerned, one has
to take into account also
the one-loop correction to the lowest order cross section,
a contribution formally proportional to $\delta(q_T^2)$.
The interference between the one-loop renormalized amplitude
with the lowest order one depends of course on the process.
Nevertheless, its singular structure is universal and allows to write
in general \cite{poli}
\begin{equation}
\label{1loop}
\cm^{(0)\dagger}_{c\bar{c}\to F}\cm^{(1)}_{c\bar{c}\to F}+{\rm
  c.c.}=\f{\as}{2\pi}\left(\f{4\pi\mu^2}{Q^2}\right)^\ep\f{\Gamma(1-\ep)}
{\Gamma(1-2\ep)}\left(-\f{2 C_c}{\ep^2}-\f{2 \gamma_c}{\ep}+{\cal A}_c^F(\phi)\right)|\cm^{(0)}_{c\bar{c}\to F}|^2.
\end{equation}
 The {\em finite} part ${\cal  A}_c^F$  depends (in general) on
the kinematics of the final state non-coloured particles and on 
the particular process in the class (\ref{class}) we want to consider.
In the case of Drell--Yan  we have \cite{dy}:
\begin{equation}
\label{ady}
{\cal A}_q^{DY}=C_F\left(-8+\f{2}{3}\pi^2\right)\, ,
\end{equation}
whereas for Higgs production in the $m_{top}\to\infty$ limit the finite contribution
is \cite{higgs}:
\begin{equation}
\label{ahiggs}
{\cal A}_g^{H}= 5 C_A +\f{2}{3} C_A \pi^2 -3 C_F \equiv 11+2 \pi^2\, .
\end{equation}
The diagonal term in Eq.~(\ref{inteqt}) can be evaluated integrating
Eq.~(\ref{sigma1p}), from $0$ to $p_T^2$,
keeping into account the contribution in
Eq.~(\ref{1loop}) and subtracting the following factorization counterterm
in the \ms\ scheme:
\begin{equation}
\label{factorization}
R^{(FCT)}_{c{\bar c}}(N)= -\f{2}{\epsilon} \f{\Gamma(1-\ep)}{\Gamma(1-2\ep)}
\left(\f{4\pi\mu^2}{\mu^2_F}\right)^\epsilon
   \gamma_{cc}^{(1)}(N) 
\, .
\end{equation}
As for the non-diagonal contribution, one needs
$\Sigma_{i{\bar c}}(N)$,
that can be computed, analogously to Eq.~(\ref{sigma1p}) as
\begin{align}
\Sigma_{i{\bar c}}(N)&= \f{1}{\Gamma(1-\ep)}
\left(\f{4\pi \mu^2}{q_T^2}\right)^\ep
\int_0^{1-2 q_T/Q} dz\,z^N \f{(1-z)\hat
  {P}_{ci}(z,\epsilon) }{\r}\nn\\
&\to 
\f{1}{\Gamma(1-\ep)}
\left(\f{4\pi \mu^2}{q_T^2}\right)^\ep
\int_0^1 dz\,z^N \hat
  {P}_{ci}(z,\epsilon)
\end{align}
where the functions ${\hat P}_{ci}(z,\epsilon)$
are the non-diagonal AP
splitting kernels
\begin{align}
\label{pgq}
{\hat P}_{gq}(z,\ep)&=C_F\left[\f{1+(1-z)^2}{z}-\ep z\right]\\
\label{pqg}
{\hat P}_{qg}(z,\ep)&=T_R\left[1-\f{2z(1-z)}{1-\ep}\right]\, ,
\end{align}
and the absence of singularities as $z\to 1$ has been exploited
to set $q_T\to 0$ in the integral.

The factorization counterterm to be subtracted in this case is
\begin{equation}
\label{factorization1}
R^{(FCT)}_{i{\bar c}}(N)= -\f{1}{\epsilon} \f{\Gamma(1-\ep)}{\Gamma(1-2\ep)}
\left(\f{4\pi\mu^2}{\mu^2_F}\right)^\epsilon
   \gamma_{ci}^{(1)}(N) 
\, .
\end{equation}
Comparing the total results to Eqs.~(\ref{inqt})
we obtain for $C^{(1)}_{ab}$:
\begin{equation}
\label{coeff}
C^{(1)F}_{ab} (z)= -{\hat P}^{\ep}_{ab}(z) + \delta_{ab}\, \delta(1-z) \left( C_a\,
  \f{\pi^2}{6}+\f{1}{2} {\cal A}_a^F(\phi) \right) \, ,
\end{equation}
where ${\hat P}^{\ep}_{ab}(z)$ represent
the ${\cal O}(\ep)$ term in the AP
${\hat P}_{ab}(z,\ep)$ splitting kernels in Eqs.~(\ref{pqq}, \ref{pgg}, \ref{pgq}, \ref{pqg}) and are given by:
\begin{align}
\label{Peps}
{\hat P}_{qq}^\ep(z)&=- C_F\, (1-z) \nn \\
{\hat P}_{gq}^\ep(z)&=- C_F\, z \nn \\
{\hat P}_{qg}^\ep(z)&=-2 T_R\,  z (1-z) \nn \\
{\hat P}_{gg}^\ep(z)&= 0\, .
\end{align}
As can be observed, the coefficient function contains both a {\it hard} process dependent contribution (proportional to ${\cal A}_a^F(\phi)$) originated in
 the one-loop correction  as well as a {\it `residual' collinear} contribution 
proportional the $\ep$ part of the splitting functions which has origin in the particularities of the \ms\ scheme (see Eq.~(\ref{factorization})), where only the $\ep=0$ (and not the full) component of the splitting functions is factorized.
The general expression in Eq.~(\ref{coeff}) reproduces correctly the
coefficient $C_{ab}^{(1)}$ computed for Drell--Yan \cite{ds}, Higgs production in the
$m_{top}\to\infty$ limit \cite{yuan,kauff}, $\gamma\gamma$ \cite{2gamma} and $ZZ$ \cite{ZZ} production.

Summarizing the {${\cal O}(\as)$ results,  the coefficients $A_c^{(1)}$ and $B_c^{(1)}$ are fully
determined by the {\it universal} 
properties of soft and collinear emission. 
The function $C_{ab}^{(1)}$ depends instead on the process through
the one-loop corrections to the LO matrix element.

\section{The calculation at ${\cal O}(\as^2)$: the quark channel}

At \oass\ $\Sigma_{q\qb}(N)$ receives two contributions:
\begin{itemize}
\item Real emission of two partons recoiling against the final state system $F(Q^2,\qt^2,\phi)$;
\item Virtual corrections to single-gluon emission.
\end{itemize}

In the following we compute these contributions in turn.

\label{sec:qoas2}

\subsection{Real corrections}
\label{qsecreal}

The computation of the double real corrections to $\Sigma_{q{\bar q}}(N)$ represents the most involved part of the complete calculation.
The difficulties arise both from the fact that the
additional 
parton in the final state implies
three more phase space 
integrals, and from the appearence of many more singular configurations that contribute to the limit $q_T \to 0$.

The kinematics for the double real emission process
$c\bar{c}\to i+j+F$ is (see Fig.~\ref{fig:dreal})
\begin{equation}
p_1+p_2\to k_1+k_2+q\, ,
\end{equation}

\begin{figure}[htb]
\begin{center}
\begin{tabular}{c}
\epsfxsize=6truecm
\epsffile{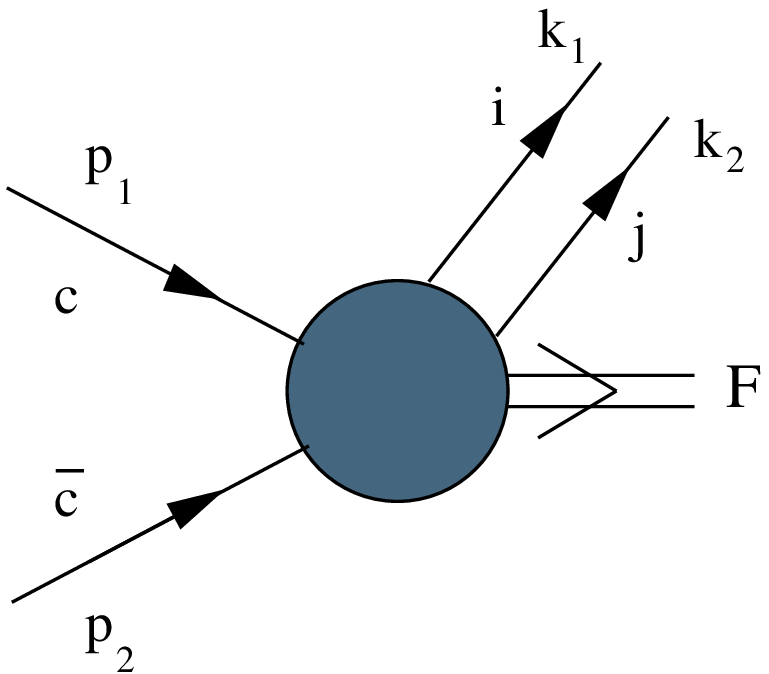}\\
\end{tabular}
\end{center}
\caption{\label{fig:dreal}{\em \oass\ contribution from double real emission.}}
\end{figure}

and the corresponding matrix element is denoted by
${\cal M}^{(0)}_{c\cb\to ij\, F}(p_1,p_2,k_1,k_2,\phi)$.
The usual invariants are defined as
\begin{equation}
s=(p_1+p_2)^2~~~~t=(p_1-q)^2~~~~u=(p_2-q)^2~~~~s_2=(k_1+k_2)^2 \, ,
\end{equation}
and fulfill the following relations
\begin{equation}
s+t+u=Q^2+s_2~~~~~~~~~~~~q_T^2=\f{ut-s_2 Q^2}{s}.
\end{equation}

In terms of these invariants, the real contribution to the cross section at fixed $q_T^2$ is given by
\begin{equation}
\label{dreal}
\f{d\sigma_{c\cb\to ij\, F}}{dq_T^2dQ^2d\phi}
=\int\f{|{\cal M}^{(0)}_{c\cb\to ij\, F}
(p_1,p_2,k_1,k_2,\phi)|^2}{2s}
\f{(s_2q_T^2)^{-\ep}}{(4\pi)^{4-2\ep}\Gamma(1-2\ep)}\f{du}{Q^2-u}\f{d\Omega}{2\pi}
\end{equation}
where $d\Omega$ is 
\begin{equation}
d\Omega=\sin \theta_2^{-2\ep}d\theta_2 \sin \theta_1^{1-2\ep}d\theta_1 \, ,
\end{equation}
with the angles defined in the frame where the partons corresponding to momentum $k_1$ and $k_2$ are back-to-back \cite{Ellis:1983hk}.

We see from Eq.~(\ref{dreal}) that the first step of the calculation involves the integration over the two angles.
The integrals needed here are typical of heavy quark production at NLO and
 most of them can be found in Refs.~\cite{angu,bojakangu}. 
The results of the angular integrals contain poles up to $1/\ep$ while terms 
that develop an extra additional singularity as $s_2\to 0$ have to be
computed up to ${\cal O}(\ep)$.

The second step is
the integration over $u$ (or $s_2$). The integration limits are given by the two roots $u_{max}$ and $u_{min}$ in Equation (\ref{umaxumin}).
At this point, it is  convenient to define the `symmetric' value for which $u=t$
\begin{equation}
u_0=Q^2-\sqrt{s(q_T^2+Q^2)} \,.
\end{equation}
This value of $u$ corresponds also to the maximum of $s_2$
\begin{equation}
s_2^{max}\equiv A= Q^2 \f{1+z-2 \sqrt{1+q_T^2/Q^2} \sqrt{z}}{z}
\end{equation}
 and, in the CM frame of $p_1$ and $p_2$, to the
configuration where $q_z=0$. 
The singularity in $s_2$ is made manifest by use of the identity
\begin{equation}
\label{reguls2}
s_2^{-1-\ep}=-\f{1}{\ep}\delta(s_2)\left(1-\ep \log A+\f{1}{2} \ep^2 \log^2 A\right)
+\f{1}{(s_2)_{A+}} -\ep\left(\f{\log s_2}{s_2}\right)_{A+}
+{\cal O}(\ep^2) \, ,
\end{equation} 
with the distributions  defined as:
\begin{equation}
\int_0^A ds_2\f{f(s_2)}{(s_2)_{A+}}=\int_0^A\f{ds_2}{s_2}(f(s_2)-f(0))
\end{equation}
\begin{equation}
\int_0^A ds_2~f(s_2)\left(\f{\log s_2}{s_2}\right)_{A+}=\int_0^A~ ds_2~\f{\log s_2}{s_2}(f(s_2)-f(0)) \, .
\end{equation}

In order to obtain $\Sigma(N)$ one finally has to integrate over $z$, keeping only
the terms that do not vanish in the small-$\qt$ limit.
At the beginning we consider only the $N=0$ moment\footnote{Notice that the calculation of a single moment is enough to obtain the resummation coefficients $A^{(2)}$ and $B^{(2)}$.}.
We will later show how to perform the calculation for general $N$, 
once one moment is known, in a simpler way.
Notice that after implementing the regularization of the $s_2=0$ singularities using Eq.~(\ref{reguls2}), the last two integrals can be performed directly in four-dimensions, since the small transverse momentum $q_T$ acts as a regulator of other possible singularities.

The double real contributions to (the non-singlet part of) $\Sigma^{(2)}_{q{\bar q}}(N)$
fall into three classes, according to the possible different final states:
\begin{itemize}
\item $q+\qb\to \qb+q+F$
\item $q+q\to q+q+F$
\item $q+\qb\to g+g+F$ \, .
\end{itemize}
Notice that the $q+q\to q+q+F$ is needed to form the non-singlet combination.

As we did at \oas, to study the small $q_T$ behaviour of $\Sigma$ we will rely on
the structure of soft and collinear singularities of the corresponding QCD matrix element.
In principle there are, of course, configurations where the two final state partons 
are hard and emitted back-to-back with small total transverse momentum. Nevertheless, 
 these configurations do not 
produce any singularities when $q_T\to 0$  and thus may be neglected.
Finally, notice that we consider only {\it double} singularities, i.e. 
configurations where  two extra partons are either collinear or soft,
 without caring about {\it single} singularities.
Configurations with only one collinear or soft parton (and the other hard)
 do not contribute to $\Sigma^{(2)}$
 since the system $F$ is not emitted with small $q_T$ in such case.

\subsubsection{Contribution from $q{\bar q}$ and $qq$ emission}
\label{qqbqq}

For the
$q+\qb\to \qb+q+F$ contribution
we have three singular regions at \oass\ \cite{Catani:2000ss}:
\begin{itemize}
\item first triple-collinear region: $k_1p_1\sim k_2 p_1\sim k_1k_2\to 0$;
\item second triple-collinear region: $k_1p_2\sim k_2 p_2\sim k_1k_2\to 0$;
\item double-soft region: $k_1,k_2\to 0$ \, .
\end{itemize}
In the first region the singularity is controlled by the following
collinear factorization formula \cite{Campbell:1998hg,Catani:1999nv,Catani:2000ss}
\begin{equation}
\label{qqbcoll}
|\cm^{(0)}_{q\qb\to q\qb F}(p_1,p_2,k_1,k_2,\phi)|^2\simeq\f{(8\pi\mu^{2\ep}\as)^2}{u^2} {\hat
  P}_{q\to \qb_1q_2(q_3)}|\cm_{q\qb\to F}^{(0)}(z_3p_1,p_2,\phi)|^2 \, ,
\end{equation}
where ${\hat P}_{q\to \qb_1q_2(q_3)}$ is the splitting function that controls 
the collinear decay of an initial state quark of momentum $p_1$ into a final state
quark-antiquark pair $\qb_1q_2$ of momenta $k_1$ and $k_2$ and the `off-shell' quark $q_3$ that participates in the hard cross section..
The explicit expression of ${\hat P}_{q\to \qb_1q_2(q_3)}$ is obtained
from the one of ${\hat P}_{\qb_1q_2q_3}$, 
the splitting function for the decay of a (`off-shell') quark into
a final state quark-antiquark pair plus a quark, given in Eq.~(\ref{qqqsf}),
with the following definitions
\begin{align}
\label{croqqb}
& s_{12}=s_2~~~s_{13}=-2p_1k_1~~~s_{23}=-2p_1k_2\nn\\
& x_1=-z_1/z_3~~~x_2=-z_2/z_3~~~x_3=1/z_3 \, ,
\end{align}
where $z_1$ and $z_2$ are the momentum fractions of
$\qb_1$ and $q_2$ ($z_3=1-z_1-z_2$).
Notice that Eq.~(\ref{croqqb})
corresponds to the following
 transformation:
\begin{equation}
r_1\to k_1,~~~~r_2\to k_2,~~~~r_3\to -p_1 \, ,
\end{equation}
applied to the expression in Eq.~(\ref{qqqsf}) to cross the `off-shell' parton to the final state.

A formula similar to Eq. (\ref{qqbcoll}) can be written in the second collinear region.

In the double-soft region the factorization formula
is instead \cite{Catani:2000ss}:
\begin{equation}
\label{qq4}
|{\cal M}^{(0)}_{q\qb\to q\qb F}\,(p_1,p_2,k_1,k_2,\phi)|^2 \simeq
(4 \pi\mu^{2\ep} \as)^2\, C_F\, T_R
\; \Big( {\cal I}_{11} +
{\cal I}_{22} -2\,{\cal I}_{12} \Big) \;
|{\cal M}^{(0)}_{q\qb \to F}\,(\phi)|^2 \, ,
\end{equation}
where
\begin{equation}
{\cal I}_{ij}={\cal I}_{ij}(k_1,k_2)= \f{p_i k_1\; p_j k_2
+ p_j k_1\; p_i k_2 - p_i  p_j 
\;k_1 k_2}{(k_1  k_2)^2 
\,p_i(k_1+k_2)\; p_j (k_1+k_2)}\, .
\end{equation}
The reader can easily check that by defining the momentum fractions
in Eq.~(\ref{croqqb})
as\footnote{To parameterize the triple-collinear limit it is necessary to introduce an additional light-cone vector $n$. This definition corresponds to the choice $n=p_2$. Notice that a similar definition can be adopted also at \oas\ to reobtain  Eq.~(\ref{qtfact})  in the small $q_T$ limit.}
\begin{equation}
\label{qqbz}
z_1=\f{k_1p_2}{p_1p_2}~~~~~z_2=\f{k_2p_2}{p_1p_2}
\end{equation}
Eq.~(\ref{qqbcoll}) correctly keeps into account also the double-soft limit
in Eq.~(\ref{qq4}).
Thus, at least outside the second collinear region, the factorization
formula (\ref{qqbcoll}) with the definitions (\ref{qqbz})
correctly gives the full singular behaviour in this channel.

The strategy to perform the calculation is the following. We use Eq.~(\ref{qqbcoll}) to approximate
the matrix element in its region of validity 
 and compute
 its contribution to $\Sigma^{(2)R}_{q\qb}(0)$ by integrating only 
in half of the phase space, that is from $u_0$ to $u_{max}$. 
The remaining region, which is obtained by exchanging $u\lra t$, will give
, due to the symmetry of the initial state, exactly the same contribution and it is taken into account
 by multiplying the computed result by 2.
As it happens at leading order, the information on the process, embodied in
the Born matrix element is completely
factored out in the calculation and disappears in $\Sigma$.
In fact the Born matrix element can be fully written in terms of the (fixed)
kinematics of the final state particles  $|\cm^{(0)}_{q\qb\to F}(z_3 p_1,p_2,\phi)|^2\equiv 
|\cm_{q\qb\to F}^{(0)}(\phi)|^2$.

For the $q+q\to q+q+F$ contribution, needed to form the non-singlet contribution in 
Eq.~(\ref{eq:nonsing}), 
 there are only two singular configurations:
\begin{itemize}
\item first triple-collinear region: $k_1p_1\sim k_2 p_1\sim k_1k_2\to 0$;
\item second triple-collinear region: $k_1p_2\sim k_2 p_2\sim k_1k_2\to 0$ \, .
\end{itemize}

For the first collinear region we can write:
\begin{equation}
\label{qqcoll}
|\cm^{(0)}_{qq\to qq F}(p_1,p_2,k_1,k_2,\phi)|^2\simeq\f{(8\pi\mu^{2\ep}\as)^2}{u^2} 
{\hat P}_{q\to q_1q_2(\qb_3)}|\cm^{(0)}_{q\qb\to F}(p_1,z_3 p_2,\phi)|^2 \,. 
\end{equation}
Here ${\hat P}_{q\to q_1q_2(\qb_3)}$ is now the splitting function which controls 
the collinear decay of an initial state quark into a final state $qq$
pair. The explicit expression for ${\hat P}_{q\to q_1q_2(\qb_3)}$ can be
obtained from the expression of ${\hat P}_{\qb_1q_2q_3}$
in Eq.~(\ref{qqqsf})
with the following definitions
\begin{align}
& s_{12}=-2p_1k_2~~~s_{13}=-2p_1k_1~~~s_{23}=s_2\nn\\
& x_1=1/z_3~~~x_2=-z_2/z_3~~~x_3=-z_1/z_3 \, ,
\end{align}
i.e., corresponding to the crossing transformation:
\begin{equation}
r_1\to -p_1,~~~~r_2\to k_2,~~~~r_3\to k_1\, ,
\end{equation} 
and similarly for the second collinear configuration (with $p_1\leftrightarrow p_2$).
There is a partial cancellation between the $C_FT_R$
contribution to $\Sigma_{q{\bar q}}^{(2)}$ from Eqs.~(\ref{qqbcoll}) and (\ref{qqcoll}), due to
the non-singlet combination.
Once this cancellation is carried out, the part corresponding to the production of 
`non-identical' partons in the $q\qb$ channel gives the following
$C_FT_R$ contribution to $\Sigma_{q{\bar q}}^{(2)}$ :
\begin{align}
\label{risqbarqnid}
\Sigma^{(2)R}_{q{\bar q}\,{\rm (nid)}}
(0)=C_F n_f T_R \,\K\Bigg[ -\f{2}{3}\f{1}{\ep}\funpq 
-\f{4}{3}\log^2\f{Q^2}{q_T^2}-\f{2}{9}\log\f{Q^2}{q_T^2}+1+\f{2}{9}\pi^2\Bigg] \, ,
\end{align}
where
\begin{equation}
\label{kappa}
\K=\f{1}{\Gamma(1-2\ep)} \left(\f{4\pi\mu^2}{q_T^2}\right)^\epsilon
\left(\f{4\pi\mu^2}{Q^2}\right)^\epsilon \,,
\end{equation}
and the explicit expression of function $\funpq$, defined in Eq.~(\ref{sigma1p}) is
\begin{equation}
\funpq=2\log \f{Q^2}{\qt^2}-3-\ep\,.
\end{equation}
At the beginning of Eq.~(\ref{risqbarqnid}) we have isolated a divergent term which will be  cancelled by a similar one appearing in the virtual contribution.

The part corresponding to the production of 
`identical' partons in the $q\qb$ channel gives also
a $C_F(C_F-C_A/2)$ contribution, which does not contain any $\log Q^2/q_T^2$
term.
Therefore, there is a great simplification in
the calculation since 
 $q_T$ can be set to zero just after performing the angular integrations.
We find:
\begin{equation}
\label{risqbarqid}
\Sigma^{(2)R}_{q\qb\,{\rm (id)}}(0)=C_F(C_F-\f{1}{2} C_A) \Big(-6+2\pi^2-16 \zeta(3)\Big) \, .
\end{equation}
The calculation of the $qq$ contribution can be performed with exactly the same strategy as for the
$q\qb$ channel\footnote{A factor $1/2$ has been included to account for
 the two  identical particles in the final state.}.
After the $C_FT_R$ contribution has been cancelled with a
similar one in the $q\qb$ channel only a   
contribution proportional
to $C_F(C_F-C_A/2)$ remains\footnote{The overall minus sign here is due to the fact that this quantity must be subtracted in order to construct the non-singlet combination.}
\begin{equation}
\label{risqq}
\Sigma^{(2)R}_{q{\bar q}({\rm qq})}(0)=-C_F(C_F-\f{1}{2} C_A) \left(\f{13}{2}-\pi^2+4
  \zeta(3)\right) \, .
\end{equation}

\subsubsection{Contribution from $gg$ emission}
\label{gg}

The calculation of the double-gluon emission correction to $\Sigma_R^{(2)}$ is more difficult, because it is not possible to keep into account all  possible singular
 configurations by using only the triple-collinear splitting functions.
We will divide the calculation in two parts, according to the corresponding colour factors. First we will consider the non-abelian, $C_FC_A$ term, which turns out to be simpler, and finally the abelian, $C_F^2$ part.

\noindent \underline{$C_FC_A$ contribution}

For this colour structure there are three singular regions to be considered \cite{Catani:2000ss}

\begin{itemize}
\item first triple-collinear region: $k_1p_1\sim k_2 p_1\sim k_1k_2\to 0$;
\item second triple-collinear region:  $k_1p_2\sim k_2 p_2\sim k_1k_2\to 0$;
\item double-soft region: $k_1,k_2\to 0$ \, .
\end{itemize}

We point out that, as discussed in Ref.~\cite{Catani:2000ss}, thanks to the coherence properties of soft-gluon radiation, the soft-collinear region does not give any contribution proportional to $C_FC_A$ (see later).

In the first collinear region the singularity is controlled by the following factorization formula:
\begin{equation}
\label{ggnabcoll}
|\cm^{(0)}_{q\qb\to gg\,F}(p_1,p_2,k_1,k_2,\phi)|_{{\rm
    nab}}^2\simeq\f{(8\pi\mu^{2\ep}\as)^2}{u^2} C_F C_A {\hat
  P}^{{\rm (nab)}}_{q\to g_1g_2(q_3)}|\cm^{(0)}_{q\qb\to F}(z_3p_1,p_2,\phi)|^2 \, ,
\end{equation}
where ${\hat P}^{{\rm (nab)}}_{q\to g_1g_2(q_3)}$ is the non-abelian part of the splitting function that controls the collinear decay of an initial state quark into a final state gluon pair. This function can be obtained
from Eq.~(\ref{qggnabsf})
with the replacement in Eq.~(\ref{croqqb}).

A  similar formula to Eq.~(\ref{ggnabcoll}) can be written in the second
collinear region (by $p_1\leftrightarrow p_2$ exchange).

In the double-soft region the factorization formula
is instead (see Eq.~(A.3) of Ref.~\cite{Catani:2000ss}):
\begin{align}
\label{dsoftnab}
|\cm^{(0)}_{q\qb\to ggF}(p_1,p_2,k_1,k_2,\phi)|_{{\rm
    nab}}^2 &\simeq
(4 \pi \mu^{2\ep} \as)^2\, C_F\, C_A\nn\\
&\cdot
\left(
2\,{\cal S}_{12}(k_1,k_2) - {\cal S}_{11}(k_1,k_2)
-{\cal S}_{22}(k_1,k_2) \right)
 |{\cal M}_{q\qb\to F}\,(\phi)|^2 \, ,
\end{align}
where the non-abelian double-soft function reads
\begin{align}
{\cal S}_{ij}(k_1,k_2) &= \f{(1-\ep)}{(k_1  k_2 )^2} \;
\f{p_i  k_1 \;p_j  k_2 + p_i  k_2 \;p_j  k_1}
{p_i (k_1+k_2) \; p_j (k_1+k_2)} \nn \\
\label{dsoftfun}
&- \f{(p_i  p_j)^2}{2 p_i k_1 \; p_j k_2 \;
p_i k_2 \; p_j k_1}
\left[ 2 - \f{p_i  k_1 \;p_j  k_2 + p_i  k_2 \;p_j k_1}
{p_i (k_1+k_2) \; p_j (k_1+k_2)} \right] \\
&+ \f{p_i p_j}{2 k_1  k_2}
\left[ \f{2}{p_i k_1 \;p_j  k_2} + 
       \f{2}{p_j k_1 \;p_i  k_2} \right. \nn \\
&- \left.
       \f{1}{p_i (k_1+k_2) \; p_j (k_1+k_2)}
   \left( 4 + 
  \f{(p_i  k_1 \;p_j  k_2 + p_i  k_2 
  \;p_j  k_1)^2}{p_i k_1 \; p_j k_2 \;
  p_i k_2 \; p_j k_1}
\right) \right] \;\;. \nn
\end{align}

As it happens in the $q\qb$ and $qq$ channels, it turns out that by defining the momentum fractions of the gluons as  in Eq.~(\ref{qqbz}), the factorization formula in  Eq.~(\ref{ggnabcoll}) correctly accounts also for the double-soft configuration.
 Furthermore, we have verified that Eq.~(\ref{ggnabcoll})
 does not introduce any additional spurious singularities in the other infrared 
configurations. Thus for this colour structure the situation is similar to the one in 
the $q\qb$ and $qq$ channels and we can follow the same strategy.
We approximate the non-abelian part of
the matrix element in the region from $u_0$ to $u_{max}$ using
Eq.~(\ref{ggnabcoll}).  We first perform the angular integrals and then,
 exploiting the
$p_1\lra p_2$ symmetry, do the remaining $u$  and $z$ integrations only over 
half of the phase space, i.e. with $u$ from $u_0$ to $u_{max}$.

In order to perform the last two steps, that are considerably more complicated than in the case of $q{\bar q}$ emission, we developed {\tt Mathematica} \cite{mathematica} 
programs
that are able to handle the cumbersome intermediate expressions in the small $\qt$ limit.

The result is\footnote{A factor $1/2$ has been included
to account for the two 
identical particles in the final state.}
\begin{align}
\label{risggnab}
\Sigma^{(2)R}_{q{\bar q}({\rm ggnab})}(0)
&=C_FC_A\,\K\Bigg[\left(\f{1}{\ep^2}+\f{1}{\ep}\left(\f{11}{6}+\log\f{Q^2}{q_T^2}\right)\right)\funpq \nn\\
&+\log^3\f{Q^2}{q_T^2}+\f{13}{6}\log^2\f{Q^2}{q_T^2}+\left(\f{35}{18}-\f{2}{3}\pi^2\right)\log\f{Q^2}{q_T^2}+4\zeta(3)-2+\f{7}{18}\pi^2\Bigg] \, ,
\end{align}
in agreement with Ref.~\cite{davies}. The first line of Eq.~(\ref{risggnab})
comes from the singular $\delta(s_2)$ terms and will be exactly cancelled by a 
contribution appearing in the virtual correction.

\noindent \underline{$C_F^2$ contribution}

For this colour structure there are six singular regions (plus the ones generated from permutations like $k_1 \leftrightarrow k_2$)  to be considered \cite{Campbell:1998hg,Catani:2000ss}

\begin{itemize}
\item first triple-collinear region: $k_1p_1\sim k_2 p_1\sim k_1k_2\to 0$;
\item second triple-collinear region:  $k_1p_2\sim k_2 p_2\sim k_1k_2\to 0$;
\item double-soft region: $k_1,k_2\to 0$;
\item first soft-collinear region: $k_1\to 0$, $k_2 p_1\to 0$;
\item second soft-collinear region: $k_1\to 0$, $k_2 p_2\to 0$;
\item double-collinear region: $k_1p_1\to 0$, $k_2 p_2\to 0$.

\end{itemize}
In the first region the singularity is controlled by the collinear factorization formula:
\begin{equation}
\label{ggabcoll}
|\cm_{q\qb\to gg\,F}(p_1,p_2,k_1,k_2,\phi)|_{{\rm
    ab}}^2\simeq\f{(8\pi\mu^{2\ep}\as)^2}{u^2} C_F^2{\hat P}^{{\rm (ab)}}_{q\to g_1g_2(q_3)}|\cm^{(0)}_{q\qb\to F}(z_3p_1,p_2,\phi)|^2 \, ,
\end{equation}
where ${\hat P}^{{\rm (ab)}}_{q\to g_1g_2(q_3)}$ is the abelian part of the
splitting function that controls the collinear decay of an initial state quark
into a final state gluon pair.
This function can be obtained from
Eq.~(\ref{qggabsf})
with the replacement in Eq.~(\ref{croqqb}). 

In the double-soft region the factorization formula is
obtained by factorizing the two eikonal factors
for independent gluon emissions
(see Eq.~(A.3) of Ref.~\cite{Catani:2000ss}):
\begin{equation}
\label{dsoftab}
|\cm^{(0)}_{q\qb\to ggF}(p_1,p_2,k_1,k_2,\phi)|_{{\rm
    ab}}^2\simeq
(4 \pi \mu^{2\ep} \as)^2\, 16C_F^2\,
{\cal S}_{12}(k_1){\cal S}_{12}(k_2)
|\cm^{(0)}_{q\qb\to F}(\phi)|^2
\, ,
\end{equation}
with ${\cal S}_{12}(k)$ defined in Eq.~(\ref{eik}).

In the soft-collinear region, say when $k_2\to 0$ and $k_1p_1\to 0$
we have instead (see Eq.~(A.5) of \cite{Catani:2000ss}):
\begin{align}
\label{sc4}
|{\cal M}^{(0)}_{q\qb\to ggF}(p_1,p_2,k_1,k_2,\phi)|^2
&\simeq (4 \pi \mu^{2\ep} \as)^2
\f{2C_F}{(1-z_1)\,p_1k_1} 
\f{(p_1-k_1)p_2}{(p_1-k_1)k_2\, p_2k_2}
\,{\hat P}_{qq}(1-z_1,\ep)\nn\\
&\cdot |\cm^{(0)}_{q\qb\to F}\left((1-z_1)p_1,p_2,\phi\right)|^2 \, ,
\end{align}
where $z_1$ is the momentum fraction of the collinear gluon of momentum $k_1$
and can be identified with the one parametrizing the triple collinear splitting
in Eq.~(\ref{ggabcoll}).
Notice that, since the soft gluon of momentum $k_2$ does not
resolve the pair of collinear partons, there is no non-abelian contribution
in Eq.~(\ref{sc4}).

In the double-collinear region we have, when e.g. $k_1p_1\to 0$ and $k_2p_2\to 0$:
\begin{align}
\label{dcoll}
|{\cal M}^{(0)}_{q\qb\to ggF}(p_1,p_2,k_1,k_2,\phi)|^2 &\simeq
\f{(4 \pi \mu^{2\ep} \as)^2}{(1-z_1)p_1k_1\, (1-\zb_2)p_2k_2}
\,{\hat  P}_{qq}\left(1-z_1,\ep\right){\hat
  P}_{qq}\left(1-\zb_2,\ep\right)\nn\\
&\cdot
|{\cal M}^{(0)}_{q\qb\to F}\left((1-z_1) p_1,(1-\zb_2) p_2,\phi\right)|^2 \, ,
\end{align}
where $z_1$ and $\zb_2$ here represent the momentum fractions (see below) involved in the two
collinear splittings.

As it happens for the $C_FC_A$ contribution, Eq.~(\ref{ggabcoll}) supplemented with the definitions (\ref{qqbz}) is able to approximate correctly also the double-soft and soft-collinear regions in half of the phase space.
But, at variance with the $C_F C_A$ case,  the same formula cannot describe
correctly the double-collinear region, since that one corresponds to the
emission of gluons from different legs, i.e., with a kinematical configuration completely different from the triple-collinear case. Therefore, the strategy followed for the
other colour factors does not work in this case.

In order to overcome this problem there are in principle two strategies.
 The first one is  to split the phase space in order to isolate the
 double-collinear region and  perform the calculation separately for its contribution using the expression in Eq.~(\ref{dcoll}).
The second is to  modify  Eq.~(\ref{ggabcoll}) in order to enforce the correct
 singular behaviour in all possible limits.
 We decided to follow the second strategy and for that  we
have first studied Eq.~(\ref{ggabcoll}) with the definitions (\ref{qqbz}) and 
isolated the terms that do, incorrectly, contribute (terms with $x_2$ in the denominator)
 when the collinear gluons are emitted from the different legs.
 In this way,  we were able to find
a slight modification of ${\hat P}^{{\rm (ab)}}_{ g_1g_2q_3}$
in Eq.~(\ref{qggabsf})
that
allows to take into account the double-collinear region without spoiling the
behaviour in the other regions as
\begin{align}
\label{pabmod}
 {\hat D}_{g_1 g_2 q_3}^{{\rm (ab)}}
&=\Biggl\{\f{s_{123}^2}{2s_{13}s_{23}}
x_3\left[\left(\f{1+x_3^2}{x_1x_2}-\ep\f{x_1^2+x_2^2}{x_1x_2}\right)f_q(\zb_1)f_q(\zb_2)-\ep(1+\ep)\right]\nn\\
&+\f{s_{123}}{s_{13}}\Biggl[\left(\f{x_3(1-x_1)+(1-x_2)^3}{x_1x_2}-\ep (x_1^2+x_1x_2+x_2^2)\f{1-x_2}{x_1x_2}\right)f_q(\zb_2)+\ep^2(1+x_3)
\Biggr]\nn\\
&+(1-\ep)\left[\ep-(1-\ep)\f{s_{23}}{s_{13}}\right]
\Biggr\}+(1\lra 2) \;\;.
\end{align}
With respect to the expression of $\Ph_{g_1 g_2 q_3}^{\rm (ab)}$
of
Eq.~(\ref{qggabsf}),
the only difference is due to the introduction of the extra factors $f_q(z)$.
The function $f_a(z)$, anticipating that a similar approach will be followed in the gluonic channel, is defined by
\begin{equation}
\label{fz}
f_a(z)=\f{z}{2C_a(1-z)}\Ph_{aa}(1-z,\ep)~~~~~a=q,g\, ,
\end{equation}
where $\Ph_{aa}$ are the collinear splitting kernels
in Eqs.~(\ref{pqq},\ref{pgg}).

The function ${\hat D}_{g_1 g_2 q_3}^{{\rm (ab)}}$ depends on the new momentum
fractions $\zb_1$ and $\zb_2$ of the gluons with respect to the 
incoming antiquark of momentum $p_2$.
These
momentum fractions should be the ones relevant for the double-collinear limit. 
Our improved factorization formula is, outside the second triple-collinear region given by
\begin{equation}
\label{impro}
|\cm^{(0)}_{q\qb\to gg\, F}(p_1,p_2,k_1,k_2,\phi)|_{{\rm
    ab}}^2\simeq\f{(8\pi\mu^{2\ep}\as)^2}{u^2} C_F^2{\hat
  D}^{{\rm (ab)}}_{q\to g_1g_2(q_3)}|\cm^{(0)}_{q\qb\to F}(z_3p_1,p_2,\phi)|^2 \, ,
\end{equation}
where ${\hat D}^{{\rm (ab)}}_{q\to g_1g_2(q_3)}$ is obtained from
${\hat D}_{g_1 g_2 q_3}^{{\rm (ab)}}$ in Eq.~(\ref{pabmod}) 
with the definitions in Eqs.~(\ref{croqqb}),(\ref{qqbz})
and by setting
\begin{equation}
\label{newmom}
\zb_1=\f{p_1k_1}{p_1p_2}~~~~~~~~\zb_2=\f{p_1k_2}{p_1p_2}\, .
\end{equation}
With Eq.~(\ref{impro}) we can consistently approximate the relevant matrix element in
the region from $u_0$ to $u_{max}$ as we did in the other channels, keeping
into account all the singular regions.
In fact in the triple-collinear region $\zb_1,\zb_2\to 0$ and
$f_q(\zb_1),f_q(\zb_2)\to 1$.
Therefore, in this limit Eq.~(\ref{impro}) reduces to Eq.~(\ref{ggabcoll}).
The factors $f_q(z)$ become
relevant in the double-collinear region, since they ensure that the correct
limit is recovered when $p_1k_1\to 0$ and $p_2k_2\to 0$ (and the same for
$k_1\lra k_2$).

Notice that the  modification of the triple-collinear formula does not spoil the process independence of our calculation:
it just allows  to write an `improved' formula  that correctly interpolates
 all possible (double-) collinear and soft singularities in the region of phase
space where we have to integrate it.
Therefore, with this approach we can avoid to split the 
phase space in regions where different approximations should be applied.

It is worth noticing that the modification in Eq.~(\ref{pabmod}) makes 
the calculation more involved already at the level of the angular integrals, mostly due to the introduction of the `new' momentum fractions $\zb_1$ and 
$\zb_2$.

For this colour structure we have to subtract the contribution from the
factorization counterterm, which can be written as
\begin{eqnarray}
\f{d\sigma_{FCT}}{dq_T^2dQ^2d\phi}&=&\f{\as}{2\pi} \int\f{dx}{x} R(x,\mu_F^2)
 \left(\f{d\sigma(p_1,p_2,\phi,k)}{dq_T^2dQ^2d\phi}
 \right)_{p_1\rightarrow x p_1} \nn \\
 &+& \f{\as}{2\pi}\int\f{dx}{x} R(x,\mu_F^2)
 \left(\f{d\sigma(p_1,p_2,\phi,k)}{dq_T^2dQ^2d\phi}
 \right)_{p_2\rightarrow x p_2} \, ,
\end{eqnarray}
where ${d\sigma(p_1,p_2,\phi,k)}$ corresponds to
 the cross section for the production of $F$ and only one extra gluon
 (see Eq.~(\ref{dsigma1})) and 
\begin{equation}
R(x,\mu_F^2)=-\f{1}{\epsilon} P_{qq}^{AP}(x) \f{\Gamma(1-\epsilon)}{\Gamma(1-2
\epsilon)} \left(  \f{4\pi \mu^2}{\mu_F^2}  \right)^\epsilon \, ,
\end{equation}
with $P_{qq}^{AP}(z)= \left(\hat{P}_{qq}(z,0) \right)_+ = C_F \left( \f{1+z^2}{1-z} \right)_+$ the regularized
AP splitting function and $\mu_F$ the factorization scale.
In the limit of small $q_T$ and after taking moments with respect to $z$, the
contribution from the counterterm factorizes as
\begin{equation}
\label{fctq}
\Sigma^{(2)}_{q{\bar q}(FCT)}(N)=  2\,C_F\funqn
\left[-\f{1}{\epsilon} \K
  \left(\f{Q^2}{\mu_F^2}\right)^\epsilon  \gamma_{qq}^{(1)}(N) \right] \, .
\end{equation}
where $\funqn$ and $\K$ are defined in Eqs.~(\ref{sigma1p}) and (\ref{kappa}),
respectively.
Therefore, in the $q_T\rightarrow 0$ limit also the contribution of the
factorization counterterm becomes process independent.

Our final result for the $N=0$ moment of the factorized contribution
$\tilde{\Sigma}^{(2)R}_{q{\bar q}({\rm ggab})}(0)\equiv  \Sigma^{(2)R}_{q{\bar q}({\rm ggab})}(0)
-\Sigma^{(2)}_{q{\bar q}(FCT)}(0)$ is:
\begin{align}
\label{risggab}
\tilde{\Sigma}^{(2)R}_{q{\bar q}({\rm ggab})}(0)& =
C_F\,\K\Bigg[\left(\f{2}{\ep^2}+\f{3}{\ep}\right) 
C_F\,\funpq+\f{2}{\ep}\int_0^1 2\Ph_{qq}(z,\ep)\log z\Bigg]\nn\\
&+C_F^2\,\Bigg[-2\log^3\f{Q^2}{q_T^2}+9\log^2\f{Q^2}{q_T^2}
-\left(2+\f{2}{3}\pi^2\right)\log\f{Q^2}{q_T^2}+16\zeta(3)-\pi^2-\f{43}{4} \Bigg] \, ,
\end{align}
in agreement with the result of Ref.~\cite{davies} for Drell--Yan.
Notice that since $\gamma_{qq}^{(1)}(0)=0$ there is no contribution from the
factorization counterterm to $\hat{\Sigma}^{(2)R}_{q{\bar q}(\rm{ggab})}(0)$.
As we did for the other colour factors, we have isolated in the first line of
Eq.~(\ref{risggab}) the part that will be cancelled by a similar term in the
virtual contribution.

A comment to the results obtained so far is in order.
The formulae in Eqs.~(\ref{risqbarqnid}), (\ref{risqbarqid}),
(\ref{risqq}), (\ref{risggnab}), (\ref{risggab})
show that the contribution to $\Sigma_{q\qb}(0)$ from double real emission are actually independent on the specific process in (\ref{class}).
This feature of the double real emission, which is due to the universality of soft and collinear radiation,
will persist also in the gluon channel.
The explicit results obtained so far all agree with the ones obtained
for Drell--Yan in Ref.~\cite{davies}.

\subsection{Virtual corrections}
\label{qsecvirt}

The second part on the calculation of $\Sigma_{q{\bar q}}(N)$ 
involves the (one-loop) virtual
 corrections to single-gluon emission. The corresponding soft and collinear limits have been recently studied in 
Refs.~\cite{Bern:1998sc,Kosower:1999rx,Catani:2000pi}.
The kinematics is the same as at \oas\ and
the singularities  originated  by the same configurations discussed
above Eq.~(\ref{col1}).
In the first collinear region the interference between the tree-level
and one-loop contributions to single gluon emission
behaves as \cite{Bern:1998sc,Kosower:1999rx}
\begin{align}
\label{1loopfac}
{\cal M}^{(0)\dagger}_{q\qb\to g\, F}&\left(p_1,p_2,k,\phi\right){\cal M}_{q\qb\to g\, F}^{(1)u}\left(p_1,p_2,k,\phi\right) + {\rm c.c.} \simeq \f{4\pi\as\mu^{2\ep}}{z_1p_1k}\nn\\
\times\Big[&{\hat P}_{qq}(z_1,\ep) 
\left({\cal M}^{(0)\dagger}_{q\qb\to F}\left(z_1p_1,p_2,\phi\right){\cal M}_{q\qb\to F}^{(1)}\left(z_1p_1,p_2,\phi\right) +{\rm c.c.} \right)\nn \\
&+2g_S^2\left(\f{4\pi\mu^2}{2p_1k}\right)^\ep{\hat P}^{(1)}_{q\to (q)g}(z_1,\ep)|{\cal M}^{(0)}_{q\qb\to F}\left(z_1p_1,p_2,\phi\right)|^2\Big]\, .
\end{align}
In Eq.~(\ref{1loopfac}) there are two terms. In the first one the tree-level
 splitting kernel ${\hat P}_{qq}(z_1,\ep)$ is factorized with respect to 
the interference of the {\it renormalized} one-loop amplitude
${\cal M}_{q\qb\to F}^{(1)}$ and the
 tree level one
${\cal M}^{(0)}_{q\qb\to F}$.

The second term contains instead the 
{\it unrenormalized} one-loop correction to the splitting kernel
${\hat P}^{(1)}_{q\to(q)g}(z_1,\ep)$ times
the Born matrix element squared. 
The function
${\hat P}^{(1)}_{q\to(q)g}(z_1,\ep)$ controls the one-loop collinear splitting of
an initial state quark into a 
final state quark with momentum fraction $z_1$, in the CDR scheme.
Its explicit expression
can be derived from the results of Ref.~\cite{Bern:1998sc,Kosower:1999rx}
and is up to ${\cal O}(\ep^0)$:
\begin{align}
 {\hat P}^{(1)}_{q\to (q)g}(x,\ep)= (1-x)^{-\ep}
 \f{C_\Gamma}{(4\pi)^2} \Bigg[ & C_A {\hat
  P}_{qq}(x,\ep) \left(-\f{1}{\ep^2} +\f{1}{2} \log^2(1-x) + \Li_2 \f{1}{1-x}
  - \Li_2\,(1-x) \right) \nn\\
+ & C_F {\hat P}_{qq}(x,\ep) \left(-\f{2}{\ep} \log(x) -2 \log(x)\log(1-x)+2
  \Li_2\,(1-x)\right) \nn\\
+ & C_F(C_F-C_A)\, x\,\Bigg]  \, ,
\end{align}
where
\begin{equation}
C_\Gamma=\f{\Gamma(1+\ep)\Gamma^2(1-\ep)}{\Gamma(1-2\ep)}\, .
\end{equation}
A factorization formula similar to Eq.~(\ref{1loopfac}) holds
when the gluon is radiated by the initial state antiquark.

Let us now consider the soft region. At one-loop order,
for a general amplitude with $n$ hard partons, 
soft factorization formulae involve colour correlations between two and three
hard momentum partons in the matrix element squared \cite{Catani:2000pi}.
Nevertheless, in the case of only two hard partons 
the soft singularity is controlled by a simpler factorization formula (see Eq.~(57) of Ref.~\cite{Catani:2000pi})
\begin{align}
\label{soft1loop}
{\cal M}^{(0)\dagger}_{q\qb\to g\, F}\left(p_1,p_2,k,\phi\right)&{\cal M}_{q\qb\to g\, F}^{(1)u}(p_1,p_2,k,\phi)+{\rm c.c.}\simeq
16\pi\as\mu^{2\ep}\;C_F\nn\\
&\cdot \Bigg(
{\cal S}_{12}(k)\left({\cal M}^{(0)\dagger}_{q\qb\to F}(\phi)
{\cal M}^{(1)}_{q\qb\to F}(\phi)+{\rm c.c.}\right)
+{\cal S}^{(1)}_{12}(k)|{\cal M}^{(0)}_{q\qb\to F}(\phi)|^2 
\Bigg)\, ,
\end{align}
where
\begin{equation}
{\cal S}_{12}^{(1)}(k)=-\f{\as}{2\pi}\,C_A\,{\cal S}_{12}(k)
\f{1}{\ep^2}\,
\f{\Gamma^4(1-\ep)\Gamma^3(1+\ep)}{\Gamma^2(1-2\ep)\Gamma(1+2\ep)}
\left[4\pi\mu^2{\cal S}_{12}(k)\right]^\ep
\end{equation}
is the {\it unrenormalized} one-loop correction to the tree-level
 eikonal factor.
Likewise at \oas\ (see Eq.~(\ref{soft})), in Eq.~(\ref{soft1loop})
colour correlations are absent, and the factorization formula is similar in structure to the collinear one in Eq.~(\ref{1loopfac}).

One can verify that, as it happens at leading order, the factorization formula 
Eq.~(\ref{1loopfac}) with $z_1=z=Q^2/s$ correctly reproduces
also the behaviour in 
the soft-region, given by Eq.~(\ref{soft1loop}).

Furthermore, by expressing $k p_1$ in terms of $q_T$ and using Lorentz invariance as we did at leading order, a single factorization formula in the 
small $q_T$ limit is obtained:
\begin{align}
\label{qtoneloop}
{\cal M}^{(0)\dagger}_{q\qb\to g\, F}\left(p_1,p_2,k,\phi\right){\cal M}^{(1)u}_{q\qb\to g\, F}&\left(p_1,p_2,k,\phi\right) + {\rm c.c.}\simeq \f{4\pi\as\mu^{2\ep}}{q_T^2}\f{2(1-z)}{z}\nn\\
\times\Bigg[&{\hat P}_{qq}(z,\ep) \left({\cal M}^{(0)\dagger}_{q\qb\to F}\left(\phi\right){\cal
  M}^{(1)}_{q\qb\to F}\left(\phi\right)+ {\rm c.c.} \right)\nn\\
&+2g_S^2\left(\f{4\pi\mu^2}{q_T^2}\right)^\ep\!\!\!(1-z)^\ep{\hat P}^{(1)}_{q\to (q)g}(z,\ep)|{\cal M}^{(0)}_{q\qb\to F}\left(\phi\right)|^2\Bigg]\, ,
\end{align}
and this formula can be used to approximate the virtual contribution in the full phase space.
The same formula 
can be obtained by defining the collinear momentum
fraction in Eq.~(\ref{1loopfac}) as:
\begin{equation}
z_1=1-\f{kp_2}{p_1p_2}\, ,
\end{equation}
and similarly when $p_1\lra p_2$.
It is important to point out that,
at variance with what happens in the double real emission contribution,
here a process-dependent 
information appears, i.e. the one-loop matrix
element ${\cal M}^{(1)}_{q\qb\to F}\left(\phi\right)$.
The most general structure of the product
$\cm^{(0)\dagger}_{q\qb\to F}\cm^{(1)}_{q\qb\to F}+{\rm
  c.c.}$
is, according to Eq.~(\ref{1loop}):
\begin{equation}
\label{1loopq}
\cm^{(0)\dagger}_{q\qb\to F}\cm^{(1)}_{q\qb\to F}\,+\,{\rm
  c.c.}=\f{\as}{2\pi}\left(\f{4\pi\mu^2}{Q^2}\right)^\ep\f{\Gamma(1-\ep)}{\Gamma(1-2\ep)}\left(-\f{2C_F}{\ep^2}-\f{3C_F}{\ep}+{\cal A}_q^F(\phi)\right)|\cm^{(0)}_{q\qb\to F}|^2.
\end{equation}
In Eq.~(\ref{1loopq})
the structure of the poles in $\ep$ is universal and fixed by the flavour
of the incoming partons, whereas, as discussed in Sect. \ref{sec:oas} the finite part is parameterized by a scalar function 
${\cal A}_q^F(\phi)$ depending on the  kinematics of the final state particles.

The contribution from the UV counterterm in the \ms\ scheme is:
\begin{equation}
\label{uvct}
\Sigma^{(2)}_{UVCT}(N)=C_F\funqn\, \K\left(\f{Q^2}{\mu^2}\right)^\ep
\left(-\f{1}{\ep}\right)\beta_0\, .
\end{equation}
By approximating our matrix element with Eq.~(\ref{qtoneloop}),
the calculation can now be performed quite easily as in Eq.~(\ref{sigma1p}).
Using Eq.~(\ref{1loopq}),
and adding the contribution of the UV counterterm in Eq.~(\ref{uvct}) we find:
\begin{align}
\label{finvirtual}
\Sigma^{(2)V}_{q{\bar q}}&(0)=C_F\,\K\Bigg\{\Bigg[-\f{1}{\ep^2}(2C_F+C_A)-\f{1}{\ep}\left(C_A\log\f{Q^2}{q_T^2}+3C_F+\beta_0\right)\Bigg]\funpq \nn\\ 
&-\f{2}{\ep}\int_0^1 2\Ph_{qq}(z,\ep)\log z 
+C_A\left(-\log^3\f{Q^2}{q_T^2}+\f{3}{2}\log^2\f{Q^2}{q_T^2}+\f{\pi^2}{3}\log\f{Q^2}{q_T^2}+\f{\pi^2}{2}-\f{15}{2}-8\zeta(3)\right)\nn\\
&+C_F\left(\left(-5+\f{4}{3}\pi^2\right)\log\f{Q^2}{q_T^2}+\f{39}{2}-2\pi^2\right)
+\left(2\log\f{Q^2}{\qt^2}-3\right){\cal A}_q^F(\phi)\nn\\
&+\beta_0\log\f{Q^2}{\mu_R^2}\left(3-2\log\f{Q^2}{q_T^2}\right)\Bigg\},
\end{align}
where $\mu_R^2$ is the renormalization scale at which $\as$ is now evaluated. The terms
involving $\funpq$ and $\Ph_{qq}(z,\ep)$
in Eq.~(\ref{finvirtual}) are the ones that cancel against the corresponding terms
in Eqs.~(\ref{risqbarqnid}), (\ref{risggnab}) and (\ref{risggab}).
In the case of Drell--Yan, by using Eq.~(\ref{ady}), our result agrees with the
one of Ref.~\cite{davies}.

\subsection{Total result for the $q{\bar q}$ channel}

After adding the real and virtual contributions in 
\begin{equation}
\Sigma^{(2)}_{q{\bar q}}(0)=
\Sigma^{(2)R}_{q{\bar q}({\rm nid})}(0) + \Sigma^{(2)R}_{q\qb({\rm id})}(0)
+ \Sigma^{(2)R}_{q\qb({\rm qq})}(0) + \Sigma^{(2)R}_{q\qb({\rm ggnab})}(0) +
\tilde{\Sigma}^{(2)R}_{q\qb(\rm{ggab})}(0) + \Sigma^{(2)V}_{q\qb}(0) 
\end{equation}
all divergent terms in $\epsilon$ cancel out and we find:
\begin{align}
\label{sigma2tot0}
 \Sigma^{(2)}_{q\qb}(0) &= \log^3\f{Q^2}{q_T^2} \left[-2 C_F^2 \right] \nn\\
 & + \log^2\f{Q^2}{q_T^2} \left[ 9 C_F^2 +2 C_F \beta_0 \right] \nn\\
 & + \log\f{Q^2}{q_T^2} \left[ C_F^2 \left( \f{2}{3} \pi^2 -7\right)
   +2 C_F{\cal A}^F_q(\phi) 
 + C_F C_A \left( \f{35}{18}-\f{\pi^2}{3} \right) 
 + C_F n_f T_R \left( -\f{2}{9} \right)
  \right] \nn\\
 &+ \left[ C_F^2 \left( -\f{15}{4}-4 \zeta(3) \right) -3C_F {\cal A}^F_q(\phi) 
 + C_F C_A \left(-\f{13}{4} -\f{11}{18} \pi^2  +  6 \zeta(3)  \right)
 \right. \nn\\
  &  \left.
      \;\;\; +\;  C_F n_f T_R \left( 1 + \f{2}{9} \pi^2  \right)
    \right]\, ,
\end{align}
where we have set again $\mu_F^2=\mu_R^2=Q^2$.
It is worth noticing that the process dependence
in Eq.~(\ref{sigma2tot0})
is fully contained in
the function ${\cal A}^F_q(\phi)$.

Once one moment (the $N=0$ in this case) has been
computed, it is quite simple to extend the calculation to a general value of
$N$ by studying the combination \cite{ds} 
\begin{equation}
\Sigma(N)-\Sigma(0)=\int dz\, \left( z^N -1\right) \f{q_T^2
  Q^2}{d\sigma_0/d\phi}\f{d\sigma}{dq_T^2 dQ^2d\phi} \,.
\end{equation}
Here, the factor $\left( z^N -1\right)$ eliminates singularities in the
integrand when $z\rightarrow 1$ and allows to set $q_T=0$ once the integral
over the variable $u$ has been done (in most of the cases it is possible to
set $q_T=0$ even before integrating over $u$). In that sense the complexity of the  calculation  is considerably reduced and the result can be expressed as
\begin{align}
\label{sigmanq}
\Sigma^{(2)}_{q\bar{q}}(N) &= \log^3\f{Q^2}{q_T^2} \left[-2 C_F^2 \right] \nn\\
&+ \log^2\f{Q^2}{q_T^2} \left[ 9 C_F^2 +2 C_F \beta_0 -6 C_F \gamma_{qq}^{(1)}(N) \right] \nn\\
&+ \log\f{Q^2}{q_T^2} \left[ C_F^2 \left( \f{2}{3} \pi^2 -7  \right)
+ C_F C_A \left( \f{35}{18}-\f{\pi^2}{3} \right) 
-\f{2}{9} C_F n_f T_R  +2 C_F {\cal A}^F_{q}(\phi)
 \right. \nn\\
 &   \hspace{1.6cm} \left.
+ \left(2 \beta_0 + 12 C_F\right) \gamma_{qq}^{(1)}(N) -4 \left(\gamma_{qq}^{(1)}(N)
\right)^2
 + 4 C_F^2 \left( \pen-\f{1}{2} \right) 
 \right] \nn\\
&+ \left[ C_F^2 \left( -\f{15}{4}-4 \zeta(3)   \right)
      + C_F C_A \left(-\f{13}{4} -\f{11}{18} \pi^2  +  6 \zeta(3)  \right)
-3 C_F {\cal A}^F_q(\phi)
 \right. \nn\\
 & \hspace{0.4cm} \left.
      + C_F n_f T_R \left( 1 + \f{2}{9} \pi^2  \right)
+ 2 \gamma_{(-)}^{(2)}(N) + 2 C_F \gamma_{qq}^{(1)}(N) \left( 
    \f{\pi^2}{3} +2 \pen
 \right) 
 \right. \nn\\
 &   \hspace{0.4cm} \left.
+2 \gamma_{qq}^{(1)}(N) {\cal A}^F_q(\phi) 
 -2 C_F (\beta_0 +3 C_F) \left( \pen-\f{1}{2} \right)
    \right]\, ,
\end{align}


where $\gamma_{(-)}^{(2)}(N)$ is the non-singlet space-like
two-loop anomalous dimension \cite{2loopns}. The extraction of the resummation coefficients for the $q{\bar q}$ channel from Eq.~(\ref{sigmanq}) will be performed, along with the corresponding one for the $gg$ channel, in Section \ref{sec:final}.

\section{The calculation at \oass: the gluon channel}

The strategy for the computation of the \oass ~contributions in the gluon
 channel is the same as the one developed for the $q\bar{q}$ case. 
In a similar way,
we first consider the double real emission contribution and then the virtual
correction.

Let us first discuss the contribution coming from the factorization 
counterterm, that will be subtracted from the real corrections in the next
 subsection. By following the same steps that
lead to Eq.~(\ref{fctq}) we obtain
\begin{align}
\label{fctg}
\Sigma_{gg(FCT)}^{(2)}(N)&=2C_A\fungn\Bigg[-\f{1}{\ep}\K\left(\f{Q^2}{\mu^2_F}\right)^\ep\gamma_{gg}^{(1)}(N)\Bigg]\nn\\
&+2C_F{\cal F}_{gq}(N,\ep)\Bigg[-\f{1}{\ep}\K\left(\f{Q^2}{\mu^2_F}\right)^\ep\gamma_{qg}^{(1)}(N)\Bigg] \, .
\end{align}
Eq.~(\ref{fctg}) contains two terms.
The first one, due
to the subtraction of one collinear gluon, is analogous to the one in Eq.~(\ref{fctq}) and  contributes to both $C_A^2$ and  $C_AT_R$ colour factors.
The second term is due to the subtraction of a quark (antiquark) collinear
 to the initial state gluons, and contributes to the $C_FT_R$ part.
The function ${\cal F}_{gq}$ 
in Eq.~(\ref{fctg}) is defined as in Eq.~(\ref{sigma1p}) by
\begin{equation}
C_F\,{\cal F}_{gq}(N,\ep)\equiv\int_0^{1-2q_T/Q}dz\, z^N\f{2(1-z){\hat P}_{gq}(z,\ep)}{\r} \; .
\end{equation}
The $q_T\to 0$ limit can be safely taken and ${\cal F}_{gq}(N,\ep)$ gives
\begin{equation}
\label{fqg}
{\cal F}_{gq}(N,\ep)\to 2\int_0^1 dz\, z^N \left(\f{(1+(1-z)^2}{z}-\ep z\right)= 2 \gamma_{gq}^{(1)}(N) -2 \ep \f{1}{N+2} \, .
\end{equation}

\subsection{Real corrections}

The contributions to $\Sigma^{(2)}_{gg}$ from double real emission fall in two classes:
\begin{itemize}
\item $g+g\to q+\qb+F$
\item $g+g\to g+g+F$ \, .
\end{itemize}


The kinematics is the same as discussed at the beginning of 
Sec.~\ref{qsecreal}. 
As we did for the quark channel, we will first perform the calculation for a fixed moment and then extend it for general $N$. Since the $N=0$ moment
is divergent for the gluon channel (see e.g. Eq~(\ref{fqg})),
we start from $N=1$. Furthermore, as it happens at LO, spin-correlations appear in the collinear decay of a gluon. Nevertheless, since the correlations cancel out after integration, we will use in the collinear factorization formulae directly the spin-averaged splitting functions.

\subsubsection{Contribution from $q{\bar q}$ emission}

For this contribution the strategy followed for the $C_FT_R$ and $C_FC_A$ terms in the quark channel
applies.
The singular regions are:
\begin{itemize}
\item first triple-collinear region: $k_1p_1\sim k_2 p_1\sim k_1k_2\to 0$;
\item second triple-collinear region:  $k_1p_2\sim k_2 p_2\sim k_1k_2\to 0$;
\item double-soft region: $k_1,k_2\to 0$
\end{itemize}
In the first triple-collinear region the factorization formula reads
\begin{equation}
\label{ggqbarqcoll}
|\cm^{(0)}_{gg\to q\qb\,F}(p_1,p_2,k_1,k_2,\phi)|^2\simeq\f{(8\pi\mu^{2\ep}\as)^2}{u^2} {\hat P}_{g\to \qb_1q_2(g_3)}|\cm^{(0)}_{gg\to F}(z_3p_1,p_2,\phi)|^2 \, ,
\end{equation}
where ${\hat P}_{g\to \qb_1q_2(g_3)}$ is the splitting function that controls the decay of an initial state gluon into a final state quark-antiquark pair and a gluon.
It can be obtained from the expression of ${\hat P}_{\qb_1q_2g_3}$ that describes the decay of an (off shell) gluon into a final state quark antiquark pair plus a gluon, given in Eq.~(\ref{gqqsf}),
with the crossing transformation (\ref{croqqb}).
In the double-soft region the factorization formula is
the same as in Eq.~(\ref{qq4}) with $C_F\to C_A$ and,
likewise in the quark channel,
the soft behaviour is correctly taken into account by
Eq.~(\ref{ggqbarqcoll}) with the definitions (\ref{qqbz}).
Therefore, we can follow the strategy successfully applied
in the quark channel to obtain
\begin{equation}
\label{risggcftr}
\tilde{\Sigma}^{(2)R}_{gg({\rm q\qb ab})}(1)=C_F n_f T_R \, \K \left( -\f{8}{27} -\f{16}{9} \log\f{Q^2}{q_T^2}
  +\f{16}{9} \log\f{Q^2}{\mu_F^2} \right)
\end{equation}
and
\begin{align}
\label{risggcatr}
\tilde{\Sigma}^{(2)R}_{gg({\rm q\qb nab})}(1)&=C_A n_f T_R  \,\K\left(-\f{2}{\ep}\funpg+ \f{25}{27} -\f{2\pi^2}{9} +\f{2}{9}
  \log\f{Q^2}{q_T^2}-\f{4}{3} \log^2\f{Q^2}{q_T^2} \right. \nn \\
 & \left. -\f{4}{3} \log\f{Q^2}{\mu_F^2}  \left(-\f{11}{3}+2\log\f{Q^2}{q_T^2}\right)\right) \, .
\end{align}
In Eqs.~(\ref{risggcftr},\ref{risggcatr}) we have already subtracted the $C_F T_R$ and $C_A T_R$ terms from the factorization counterterm (with $N=1$) in Eq.~(\ref{fctg}). Furthermore, in Eq.~(\ref{risggcatr}) we have isolated a divergent term that will be cancelled by a similar term in the virtual contribution. The explicit expression
of the function $\funpg$, defined in Eq.~(\ref{sigma1p}), reads
\begin{equation}
\funpg= -\f{11}{3}+2\log\f{Q^2}{q_T^2}\, .
\end{equation}

\subsubsection{Contribution from $gg$ emission}

The calculation of the $gg$ $C_A^2$ contribution to 
$\Sigma_{gg}^{(2)}$ parallels
the one for the $C_F^2$ part in the quark channel since the 
singular configurations have the same complicated pattern as
described above  Eq.~(\ref{ggabcoll}).
The triple-collinear region is controlled by the factorization formula
\begin{equation}
\label{ggca2coll}
|\cm^{(0)}_{gg\to gg\,F}(p_1,p_2,k_1,k_2,\phi)|^2\simeq\f{(8\pi\mu^{2\ep}\as)^2}{u^2}
{\hat P}_{g\to g_1g_2(g_3)}|\cm^{(0)}_{gg\to F}(z_3p_1,p_2,\phi)|^2 \, ,
\end{equation}
where the function ${\hat P}_{g\to g_1g_2(g_3)}$ is now obtained
by applying the crossing transformation (\ref{croqqb}) to the splitting function ${\hat P}_{g_1g_2g_3}$ that controls the collinear decay of a gluon into three final state gluons, given in Eq.~(\ref{gggsfav}).

The factorization formulae in the soft-collinear and double-collinear regions
are analogous
to Eqs.~(\ref{sc4},\ref{dcoll}), and can be obtained from them by conveniently changing the colour factors ($C_F\to C_A$) and splitting functions ($\hat{P}_{qq}\to \hat{P}_{gg}$). The factorization formula in the double-soft region
receives two contributions analogous to the ones in Eqs.~(\ref{dsoftnab},\ref{dsoftab}).
 
Likewise in the quark channel, Eq.~(\ref{ggca2coll}) with the momentum 
fractions defined as in Eq.~(\ref{qqbz}) approximates correctly, again in
 half of the phase space,  all possible infrared configurations but the
 double-collinear one. 

In order to proceed further, we use the technique developed
for the quark channel. 
As before, we first 
study the behaviour of Eq.~(\ref{ggca2coll}) with the definitions (\ref{qqbz}) in the double collinear limit
and identify the terms that do (incorrectly) contribute  in that limit.
Those terms have to be modified in order to enforce the
 correct double-collinear limit without affecting the
singular behaviour in the other regions.
The modified splitting function we obtain is:
\begin{align}
\label{pggmod}
{\hat D}_{g_1 g_2 g_3}&={\hat P}_{g_1 g_2 g_3}^{\rm non-sing}+
\Bigg\{\Bigg[{\hat P}_{g_1 g_2 g_3}^{{\rm sing}-1}+\left({\hat P}_{g_1 g_3 g_2}^{{\rm sing}-1}+
{\hat P}_{g_3 g_1 g_2}^{{\rm sing}-1}\right)f_g(\zb_2)\nn\\
&+\left({\hat P}_{g_1 g_2 g_3}^{{\rm sing}-2}+{\hat P}_{g_1 g_3 g_2}^{{\rm sing}-2}\right)f_g(\zb_2)
+{\hat P}_{g_3 g_2 g_1}^{{\rm sing}-2}f_g(\zb_1)f_g(\zb_2)\Bigg]+(1\lra 2)\Bigg\} \, .
\end{align}
The first term
\begin{align}
{\hat P}_{g_1 g_2 g_3}^{\rm non-sing}&=C_A^2\Bigg\{\f{(1-\ep)}{4s_{12}^2} t^2_{12,3}+\f{3}{4}(1-\ep)
+\f{s_{123}}{s_{12}}\left[4\f{x_1x_2-1}{1-x_3}+\f{3}{2}+\f{5}{2}x_3\right]\nn\\
&+\f{s_{123}^2}{s_{12}s_{13}}\Bigg[x_2x_3-2+\f{x_1(1+2x_1)}{2}+\f{1+2x_1(1+x_1)}{2(1-x_2)(1-x_3)}\Bigg]\Bigg\}
+(5~{\rm permutations}) \, ,
\end{align}
contains the part of
${\hat P}_{g_1g_2g_3}$
in Eq.~(\ref{gggsfav})
that does not contribute to the double-collinear limit.
Therefore, this part of the splitting function does not need any modifications.
The variable $t_{ij,k}$ is defined in Eq.~(\ref{tij}).

The second part is instead modified with the introduction
of the function $f_g(z)$, defined in Eq.~(\ref{fz}).
The functions ${\hat P}_{g_1 g_2 g_3}^{{\rm sing}-1}$ and ${\hat P}_{g_1 g_2 g_3}^{{\rm sing}-2}$ are
\begin{equation}
{\hat P}_{g_1 g_2 g_3}^{{\rm sing}-1}=C_A^2\f{s_{123}}{s_{12}}\Bigg(\f{x_1x_2-2}{x_3}
+\f{\left(1-x_3(1-x_3)\right)^2}{x_3x_1(1-x_1)}\Bigg) \, ,
\end{equation}
\begin{align}
{\hat P}_{g_1 g_2 g_3}^{{\rm sing}-2}&=C_A^2 \f{s_{123}^2}{s_{12}s_{13}}\left(\f{x_1x_2(1-x_2)(1-2x_3)}{x_3(1-x_3)}
+\f{1-2x_1(1-x_1)}{2x_2x_3}\right)\, .
\end{align}
Our improved factorization formula is therefore
\begin{equation}
\label{ggca2mod}
|\cm^{(0)}_{gg\to gg\,F}(p_1,p_2,k_1,k_2,\phi)|^2\simeq\f{(8\pi\mu^{2\ep}\as)^2}{u^2} {\hat D}_{g\to g_1g_2(g_3)}|\cm^{(0)}_{gg\to F}(z_3p_1,p_2,\phi)|^2 \, .
\end{equation}
As for the quark channel, the expression of ${\hat D}_{g\to g_1g_2(g_3)}$
is obtained from the one of ${\hat D}_{g_1g_2g_3}$
in Eq.~(\ref{pggmod})
by using Eqs.~(\ref{croqqb}), (\ref{qqbz}) and defining
$\zb_1$ and $\zb_2$ through Eq.~(\ref{newmom}).

In the triple-collinear limit $f_g(\zb_1),f_g(\zb_2)\to 1$ and
the various contributions in Eq.~(\ref{pggmod}) reconstruct the
 triple-collinear splitting function
${\hat P}_{g_1g_2g_3}$.
The role of the functions $f_g$ is again 
to enforce the correct behaviour in the 
double-collinear region.
It is worth stressing that there are in principle many ways to conveniently 
modify
the splitting function and that we have tried to find the
simplest one  that fulfills all the requirements and can be 
integrated afterwards.

The function ${\hat P}_{g_1g_2g_3}$
in Eq.~(\ref{gggsfav})
has by itself the most complicated expression among the various ${\hat P}_{a_1a_2a_3}$  because
one has to sum over six permutations.
Besides that, the modification in (\ref{pggmod}) makes
the angular integration very involved. Since many of the ensuing terms have an
additional singularity as $s_2\to 0$ some of the angular integrals
in Ref.~\cite{angu} have to be evaluated one order higher in  $\ep$.
Once the angular integrals have been performed, one has to face an additional complication: `spurious' $1/\qt^2$ and $1/\qt^4$ singularities appear in
the intermediate steps,
 which of course cancel in the final result,
 but create additional problems to take the $\qt\to 0$ limit.
The final (factorized) result is
\begin{align}
\label{risggca2}
\tilde{\Sigma}^{(2)R}_{gg({\rm gg})}(1)&=C_A\,\K\Bigg[\left(\f{3}{\ep^2}  +\f{3}{\ep}
\f{11}{6} +\f{1}{\ep} \log\f{Q^2}{q_T^2}\right)C_A\funpg+\f{2}{\ep}\int_0^1
2\,z\,\Ph_{gg}(z,\ep)\log z \Bigg]  \nn\\ 
& +C_A^2\Bigg[- \log^3\f{Q^2}{q_T^2} + \f{77}{6} \log^2\f{Q^2}{q_T^2} 
-\left(\f{82}{9} + \f{4\pi^2}{3}  \right) \log\f{Q^2}{q_T^2}
-\f{533}{27}+\f{11\pi^2}{9}+10 \zeta(3)
\Bigg]\, , \nn\\
\end{align}
where we have isolated in the first line the terms that will be cancelled by
analogous virtual contributions.
Notice that, since $\gamma_{gg}^{(1)}(1)=-\f{2}{3}n_fT_R$, there is no contribution to Eq.~(\ref{risggca2}) from the factorization counterterm
in Eq.~(\ref{fctg}).

\subsection{Virtual corrections}

We finally compute the small-$\qt$ behaviour of the virtual contribution to
$\Sigma_{gg}^{(2)}$.
The calculation parallels the one for the quark in Sec.~\ref{qsecvirt}, and the singular configurations
are the same as at leading order.
For the collinear limit, say when $k p_1\to 0$, we can write a formula similar to Eq.~(\ref{1loopfac})
\begin{align}
\label{1loopfacg}
{\cal M}^{(0)\dagger}_{gg\to g\, F} &\left(p_1,p_2,k,\phi\right){\cal M}^{(1)u}_{gg\to g\, F}
\left(p_1,p_2,k,\phi\right) +{\rm c.c.}\simeq \f{4\pi\as\mu^{2\ep}}{z_1p_1k}\nn\\
\times\Big[ &{\hat P}_{gg}(z_1,\ep) \left( {\cal M}^{(0)\dagger}_{gg\to F}\left(z_1p_1,p_2,\phi\right){\cal M}^{(1)}_{gg\to F}\left(z_1p_1,p_2,\phi\right) +{\rm c.c.}\right) \nn \\
&+2g_S^2\left(\f{4\pi\mu^2}{2p_1k}\right)^\ep{\hat P}^{(1)}_{g\to (g)g}(z_1,\ep)|{\cal M}^{(0)}_{gg\to F}\left(z_1p_1,p_2,\phi\right)|^2\Big] \, ,
\end{align}
where ${\hat P}_{gg}(z_1,\ep)$ is the tree-level splitting kernel in Eq.~(\ref{pgg}) and
${\hat P}^{(1)}_{g\to (g)g}(z_1,\ep)$ is the {\it unrenormalized} one-loop correction to the AP kernel for the collinear splitting of an initial state gluon into a 
final state gluon with momentum fraction $z_1$, in the CDR scheme.
Its explicit expression can be derived from the results
of Ref.~\cite{Bern:1998sc} and is up to ${\cal O}(\ep^0)$:
\begin{align}
 {\hat P}^{(1)}_{g\to (g)g}(x,\ep)&= (1-x)^{-\ep}
 \f{C_\Gamma C_A}{(4\pi)^2} \Bigg[  {\hat
  P}_{gg}(x,\ep) \left(-\f{1}{\ep^2} -\f{2}{\ep} \log(x) -2 \log(1-x)\log(x)+\f{\pi^2}{3} \right) \nn\\
&- \f{1}{3} (C_A-2 n_f T_R ) x\Bigg]  \, .
\end{align}
%
%

A similar formula holds when the gluon is radiated by the initial state antiquark.
In the soft region, the factorization formula is the same as in Eq.~(\ref{soft1loop})
with $C_F\to C_A$ and
one can verify that, as it happens in the quark channel,
Eq.~(\ref{1loopfacg}) with $z_1=z=Q^2/s$
correctly reproduces also the behaviour in 
the soft-region \cite{Bern:1998sc,Catani:2000pi}.
In the same way as for the quark channel we can write down
a single factorization formula in the small $q_T$ limit:
\begin{align}
\label{gtoneloop}
{\cal M}^{(0)\dagger}_{gg\to g\, F}\left(p_1,p_2,k,\phi\right){\cal M}^{(1)u}_{gg\to g\, F}&\left(p_1,p_2,k,\phi\right)+{\rm c.c.}\simeq \f{4\pi\as\mu^{2\ep}}{q_T^2}\f{2(1-z)}{z}\nn\\
\times\Bigg[&{\hat P}_{gg}(z,\ep) \left( {\cal M}^{(0)\dagger}_{gg\to F}\left(\phi\right){\cal
  M}^{(1)}_{gg\to F}\left(\phi\right)+{\rm c.c.}\right)\nn\\
&+2g_S^2\left(\f{4\pi\mu^2}{q_T^2}\right)^\ep\!\!(1-z)^\ep{\hat P}^{(1)}_{g\to (g)g}(z,\ep)|{\cal M}^{(0)}_{gg\to F}\left(\phi\right)|^2\Bigg]\, .
\end{align}
According to Eq.~(\ref{1loop})
the {\it renormalized }  amplitude ${\cal M}_{gg\to F}^{(1)}$ can be written,
up to ${\cal O}(\ep^0)$ as:
\begin{equation}
\label{1loopg}
\cm^{(0)\dagger}_{gg\to F}\cm^{(1)}_{gg\to F}\,+\,{\rm
  c.c.}=\f{\as}{2\pi}\left(\f{4\pi\mu^2}{Q^2}\right)^\ep\f{\Gamma(1-\ep)}{\Gamma(1-2\ep)}\left(-\f{2 C_A}{\ep^2}-\f{2 \beta_0}{\ep}+{\cal A}_g^F(\phi)\right)|\cm^{(0)}_{gg\to F}|^2 \, .
\end{equation}
In the case of  Higgs
production, in the  $m_H \ll m_{top}$ limit and
including also the finite
renormalization to the effective $ggH$ vertex ,
the function ${\cal A}_g^H(\phi)$
is given in Eq.~(\ref{ahiggs}) .

The contribution from the UV counterterm (in the  \ms\ scheme)
 needed to renormalize the splitting kernel ${\hat P}^{(1)}_{g\to(g)g}$ is\footnote{The total UV counterterm in the case of Higgs production
would be three times this one, but we have included part of it in
the renormalized amplitude (\ref{1loopg}).}:
\begin{equation}
\label{uvctg}
\Sigma^{(2)}_{gg\,UVCT}(N)=C_A\fungn\, \K\left(\f{Q^2}{\mu^2}\right)^\ep\left(-\f{1}{\ep}\right)\beta_0 \, .
\end{equation}
By approximating our matrix element with Eq.~(\ref{gtoneloop}), using
Eq.~(\ref{1loopg}), and adding the contribution from Eq.~(\ref{uvctg}) we find
\begin{align}
\label{finvirtualgg}
\Sigma^{(2)V}_{gg}(1)&=C_A\,\K\Bigg\{\Bigg[-\f{3}{\ep^2} C_A -\f{3}{\ep} \beta_0
-\f{1}{\ep} C_A\log\f{Q^2}{q_T^2}\Bigg] \funpg-\f{2}{\ep}\int_0^1 2\,z\,\Ph_{gg}(z,\ep)\log z   \nn\\ 
& +C_A\left(-\log^3\f{Q^2}{q_T^2}+ \f{11}{6}\log^2\f{Q^2}{q_T^2}+\left(-\f{65}{18}+\f{5\pi^2}{3}\right)\log\f{Q^2}{q_T^2}-\f{11\pi^2}{6}+\f{389}{27}-8\zeta(3)\right)\nn\\
&+\f{4}{9} n_fT_R +{\cal A}_g^F(\phi) \left(-\f{11}{3}+2\log\f{Q^2}{q_T^2}\right)
- \beta_0\log\f{Q^2}{\mu_R^2}\left(-\f{11}{3}+2\log\f{Q^2}{q_T^2}\right)\Bigg\}.
\end{align}
The terms involving $\funpg$ and $\Ph_{gg}(z,\ep)$ in Eq.~(\ref{finvirtualgg}) cancel
the corresponding divergent contributions in Eqs.~(\ref{risggcatr}) and (\ref{risggca2}).

\subsection{Total result for the gluon channel}

After adding all the contributions in
\begin{equation}
\Sigma^{(2)}_{gg}(1)=\tilde{\Sigma}^{(2)R}_{gg({\rm q\qb ab})}(1)
+\tilde{\Sigma}^{(2)R}_{gg({\rm q\qb nab})}(1)
+\tilde{\Sigma}^{(2)R}_{gg({\rm gg})}(1)
+\Sigma^{(2)V}_{gg}(1) \, ,
\end{equation}
all divergent terms cancel out and
we obtain
\begin{align}
\Sigma^{(2)}_{gg}(1)&= \log^3\f{Q^2}{q_T^2} \left[-2 C_A^2 \right] \nn\\
&+ \log^2\f{Q^2}{q_T^2} \left[ 8 C_A \beta_0 +4 C_A n_f T_R   \right] \nn\\
&+ \log\f{Q^2}{q_T^2} \left[ C_A^2 \left(- \f{229}{18}+ \f{\pi^2}{3} \right)
+\f{2}{9} C_A n_f T_R -\f{16}{9} C_F n_f T_R
  + 2 C_A {\cal A}^F_{g}(\phi)
 \right] \nn\\
&+ \left[ C_A^2 \left( 2 \zeta(3) -\f{16}{3} -\f{11}{18} \pi^2 \right) 
 - \f{8}{27} C_F n_f T_R +  C_A n_f T_R \left( \f{37}{27}- \f{2}{9} \pi^2 \right) -\f{11}{3} C_A {\cal A}^F_{g}(\phi) \right]\, , \nn \\
\end{align}
where we have set again $\mu_F^2=\mu_R^2=Q^2$.

The contribution for general $N$ can be computed as explained in the previous section for the quark channel.
The total result is:
\begin{align}
\label{sigmang}
\Sigma^{(2)}_{gg}(N) &= \log^3\f{Q^2}{q_T^2} \left[-2 C_A^2 \right] \nn\\
&+ \log^2\f{Q^2}{q_T^2} \left[ 8 C_A \beta_0 -6 C_A \gamma_{gg}^{(1)}(N) \right] \nn\\
&+ \log\f{Q^2}{q_T^2} \left[ C_A^2 \left( \f{67}{9}+ \f{\pi^2}{3} \right)
-\f{20}{9} C_A n_f T_R   + 2 C_A {\cal A}^F_{g}(\phi)
 \right. \nn\\
 &   \hspace{1.7cm} \left.
+ 2 \beta_0  \left( \gamma_{gg}^{(1)}(N)-\beta_0\right) -4 \left(
    \gamma_{gg}^{(1)}(N) -\beta_0\right)^2
-4 n_f\,\gamma_{gq}^{(1)}(N) \gamma_{qg}^{(1)}(N)
 \right] \nn\\
&+ \left[ C_A^2 \left( -\f{16}{3}+ 2 \zeta(3)\right) 
 + 2 C_F n_f T_R + \f{8}{3} C_A n_f T_R
-2 \beta_0 \left( {\cal A}^F_g(\phi) + C_A \f{\pi^2}{6} \right)
 \right. \nn \\     
&  \left. + 2 \gamma_{gg}^{(2)}(N) + 2 \gamma_{gg}^{(1)}(N) \left(  {\cal A}^F_g(\phi) +
    C_A \f{\pi^2}{3} \right) +4 C_F n_f\,\gamma_{qg}^{(1)}(N) \peng
    \right]\, .
\end{align}
Here $\gamma^{(2)}_{gg}(N)$ is the
singlet space-like (gluon-gluon) two-loop anomalous dimension \cite{2loops} 
whereas the coefficient
$\peng$ has origin on the $N$ moments of
$-\hat{P}_{gq}^\ep(z)$, see Eq.~(\ref{fqg}).

\section{Final results and discussion}
\label{sec:final}

We can now compare the results obtained in the previous sections with the second order expansion of the resummation formula in Eq.~(\ref{sigma2}).
As for the $N$-dependent contributions in Eqs.~(\ref{sigmanq}), (\ref{sigmang}), they fully agree with
the ones in Eq.~(\ref{sigma2})\footnote{We have checked that the results in the quark {\em singlet} channel are also in agreement with Eq.~(\ref{sigma2}).}. 
This agreement can be considered as a non-trivial check of the validity
of the resummation formalism, because
the expressions
in Eqs.~(\ref{sigmanq}),(\ref{sigmang}) are completely general and the process dependence is fully embodied in the functions ${\cal A}^F_c(\phi)$.
As an alternative,
given the resummation formalism
for granted, the result
in  Eqs.~(\ref{sigmanq}), (\ref{sigmang})
can be considered as an independent re-evaluation of the two-loop anomalous dimensions.

As far as the $N$-independent part is concerned, it can be used to fix
the coefficients $A^{(2)}$ and $B^{(2)}$.
By comparing the single-logarithmic contributions
in Eqs.~(\ref{sigmanq}), (\ref{sigmang}) with the one in Eq.~(\ref{sigma2})
we obtain for the coefficient $A_a^{(2)}$:
\begin{equation}
A_a^{(2)}= K A_a^{(1)}~~~~~~~~~~~a=q,g \, ,
\end{equation}
where $K$ is given in Eq.~(\ref{K}), thus confirming
the results first obtained in Ref.~\cite{Kodaira:1982nh,Catani:1988vd}.
By comparing the non logarithmic terms
we find that the coefficient $B^{(2)}$ can be expressed as well by a single formula for both channels:
\begin{equation}
\label{b2}
B_a^{(2)F}=-2\, \gamma_{a}^{(2)} + \beta_0
\left( \f{2}{3} C_a \pi^2 + {\cal A}^F_a(\phi)  \right)~~~~~~~~~~~a=q,g \, ,
\end{equation}
where 
$\gamma_{a}^{(2)}$ are the coefficients of the $\delta(1-z)$ term in the
two-loop splitting functions $P_{aa}^{(2)}(z)$ \cite{2loopns,2loops},  given by
\begin{align}
\gamma_{q}^{(2)}&= C_F^2\left( \f{3}{8}-\f{\pi^2}{2}+6 \zeta(3) \right) +
 C_F C_A \left( \f{17}{24}+ \f{11 \pi^2}{18}-3 \zeta(3)\right) 
- C_F n_f T_R\left( \f{1}{6}+ \f{2 \pi^2}{9}\right) 
\nn \\ 
\gamma_{g}^{(2)}&= C_A^2\left( \f{8}{3}+ 3 \zeta(3) \right) - C_F n_f T_R -
\f{4}{3} C_A n_f T_R\, .
\end{align}
From Eq.~(\ref{b2}) we see that $B^{(2)}$, besides
the  $-2\, \gamma_{a}^{(2)}$ term which matches the expectation from the
${\cal O}(\as)$ result, receives a {\it process-dependent} contribution controlled
by the one-loop correction to the LO amplitude (see Eq.~(\ref{1loop})).
Thus, as anticipated at the beginning, although the Sudakov form factor in Eq.~(\ref{sudakov}) is usually considered universal we find that it is actually process-dependent beyond next-to-leading logarithmic accuracy.

However, by using the general expression in Eq.~(\ref{b2}) it is possible to obtain
$B^{(2)}$ for a given process just by computing the one-loop correction to the
LO amplitude for that process.
For the Drell--Yan case, by using Eq.~(\ref{ady}), our
result for $\Sigma^{(2)}_{q\bar{q}}(N)$ agrees with the one of Ref.~\cite{ds}, confirming the coefficient $B_q^{(2)DY}$ in Eq.~(\ref{b2dy}).

In the interesting case of Higgs production in the $m_{top}\to\infty$ limit, by using
Eq.~(\ref{ahiggs}) we find\footnote{Actually, using the results of Ref.~\cite{Spira:1995rr}
for the two-loop $gg\to H$ amplitude, one can also obtain
$B_g^{(2)H}$ for arbitrary $m_{top}$.}
:
\begin{equation}
\label{b2h}
B_g^{(2)H}=C_A^2\left(\f{23}{6}+\f{22}{9}\pi^2-6\zeta(3)\right)+4C_F\,n_f\,T_R-C_A\,n_f\,T_R\left(\f{2}{3}+\f{8}{9} \pi^2\right) 
 -\f{11}{2} C_F\, C_A \, .
\end{equation}
In particular, this result allows to improve the
present accuracy of the matching
between resummed predictions \cite{Balazs:2000wv}
and fixed order calculations \cite{deFlorian:1999zd}.

Notice that in this case, the coefficient $B_g^{(2)H}$ turns out to be numerically large. Actually, for $n_f=5$ we have $B_g^{(2)H}/B_g^{(1)}\approx -14 $, whereas for Drell--Yan the same ratio leads to 
$B_q^{(2)DY}/B_q^{(1)}\approx -1.9$, i.e. about 7 times smaller than
 for Higgs production. Both the appearance of a $C_A^2$ term (compared to $C_F^2$ in the quark case) and the size of the one-loop corrections to Higgs production are the reasons for the large coefficient. Clearly, the use of  
$B_g^{(2)H}$ in the implementation of the resummation formula will have an 
important phenomenological impact \cite{talksb2}.
 Actually, one can expect that the inclusion of
$B_g^{(2)H}$, which will tend to reduce the resummed cross section, will partially compensate the increase in the normalization produced by the (also) large coefficient $C_{gg}^{(1)H}$ \cite{yuan,kauff}.

The fact that the Sudakov form factor is process-dependent
is certainly unpleasant. Usually it is called the quark or gluon form factor,
since it should be determined by the universal properties of soft and collinear emission.
With the result in Eq.~(\ref{b2}), instead  we find, for example, that the form factor for $\gamma\gamma$
production is different from the one for Drell--Yan.
Moreover, since the hard function ${\cal A}_c^F$ depends in general on the details of the
kinematics of $F$ (in case of $\gamma\gamma$ production it would depend, e.g., on the rapidities of the photons), the same happens to the coefficient $B^{(2)}$ and thus to the Sudakov form factor
in Eq.~(\ref{sudakov}).

However
the results in Eq.~(\ref{coeff}) and Eq.~(\ref{b2})
suggest a simple interpretation \cite{Catani:2001vq}.
We can see in Eq.~(\ref{coeff}) that the process-dependent
coefficients functions $C^{(1)F}_{ab} (z)$ have two contributions. The first has a {\em collinear} origin and is driven by the ${\cal O}(\ep)$ part of the ${\hat P}_{ab}(z,\ep)$ kernel
(see Eqs.~(\ref{Peps})). The second has instead a {\em hard} origin, and contains the finite part of the one-loop correction to the leading order subprocess.
As a consequence, the scale at which $\as$ should be evaluated is different for these two terms.
In the collinear contribution $\as$ should be evaluated at same scale as the parton distributions are, i.e. $b_0^2/b^2$.
By contrast, the correct scale at which $\as$ should be evaluated in the hard contribution
is the hard scale $Q^2$.

As discussed in Ref.~\cite{Catani:2001vq} this mismatch,
that affects the resummation formula in its usual form Eq.~(\ref{nonunw}), 
can be solved by introducing a new process-dependent hard function $H_c^F(\as(Q^2))$. The ensuing
resummation formula is \cite{Catani:2001vq}
\begin{align}
\label{unw}
W_{ab}^{F}(s; Q, b, \phi) =& \sum_c \int_0^1 dz_1 \int_0^1 dz_2 
\; C_{ca}(\as(b_0^2/b^2), z_1) \; C_{{\bar c}b}(\as(b_0^2/b^2), z_2)
\; \delta(Q^2 - z_1 z_2 s) \nn \\
\cdot\; & \f{d\sigma_{c{\bar c}}^{F}(Q^2, \as(Q^2),\phi)}{d\phi} \;S_c(Q,b) \;\;.
\end{align}
where
\begin{equation}
\f{d\sigma_{c{\bar c}}^{F}(Q^2, \as(Q^2),\phi)}{d\phi} = \f{d\sigma_{c{\bar c}}^{(LO) \,F}(Q^2,\phi)}{d\phi} \;
H_c^{F}(\as(Q^2), \phi) \;\;.
\end{equation}
As discussed in Ref.~\cite{Catani:2001vq}, this modification is sufficient to make the Sudakov form factor $S_c(Q,b)$ and the coefficient functions $C_{ab}(\as(b_0^2/b^2),z)$ process-independent, with  $C_{ab}$ and $H_c^F$ being dependent on the introduced `resummation-scheme'.
We point out that this modification is not only a formal improvement,
since, once a resummation scheme is fixed,
the resummation coefficients in Eq.~(\ref{unw}) are now universal
and it is enough to compute the function $H_c^{F}$ at the desired order for the
process under consideration.

Summarizing, in this paper 
we have exploited  
the current knowledge on the
infrared behaviour of tree-level and one-loop QCD amplitudes at \oass\ to compute the logarithmically-enhanced contributions
up to next-to-next-to-leading logarithmic accuracy, in an general way, for both quark and gluon channels.
Comparing our results with the $q_T$-resummation formula
we have extracted the coefficients that control the resummation of the 
large logarithmic contributions. We have presented
a result that allows to compute
 the resummation coefficient $B^{(2)F}$ for any process, by simply knowing 
the one-loop (virtual) corrections to the lowest order result.
In particular we have obtained the result for the case of Higgs production in the large $m_{top}$ approximation,
 which turns out to be numerically relevant for phenomenological analyses.

The results of our calculation clearly show that the Sudakov form factor is actually process dependent within the conventional resummation approach.
An improved version of the resummation formula 
where this problem is absent
has been presented in Ref.~\cite{Catani:2001vq}.

\section*{Acknowledgements}

We thank Stefano Catani for a fruitful collaboration and helpful discussions,
 Zoltan Kunszt, James Stirling, Luca Trentadue and Werner Vogelsang for discussions,
and Christine Davies for providing us with a copy of her PhD thesis, where the details of the calculation for Drell--Yan are shown.

This work has been almost entirely performed at 
the Institute of Theoretical Physics at the ETH-Zurich.
We thank Zoltan Kunszt and the staff of ETH for
the warm hospitality and for the pleasant time we spent in Zurich.

\renewcommand{\theequation}{A.\arabic{equation}}
\setcounter{equation}{0}

\section*{Appendix: Triple-collinear splitting functions}
\label{appendixa}

In this Appendix we collect the various expressions of the triple-collinear splitting functions.
Denoting by $r_1,\, r_2$ and $r_3$ the momenta of the final state partons that become collinear,
the triple-collinear splitting functions depend on the invariants
$s_{ij}=(r_i+r_j)^2$, $s_{123}=s_{12}+s_{13}+s_{23}$
that parameterize how the collinear limit is approached,
and on the momentum fractions 
$x_i $ ($i=1,2,3$)
 involved in the
 collinear splitting.
The splitting function for the collinear decay of a quark $q$ in $q\qb$ pair plus a quark is
\begin{equation}
\label{qqqsf}
\Ph_{{\bar q}_1q_2q_3} \, =
\left[ \Ph_{{\bar q}^\prime_1q^\prime_2q_3} \, + \,(2\lra 3) \,\right]
+  \Ph^{({\rm id})}_{{\bar q}_1q_2q_3} \;\;,
\end{equation} 
where
\begin{equation}
\label{qqqprimesf}
 \Ph_{{\bar q}^\prime_1 q^\prime_2 q_3} \, = \f{1}{2} \, 
C_F  T_R \,\f{s_{123}}{s_{12}} \left[ - \f{t_{12,3}^2}{s_{12}s_{123}}
+\f{4x_3+(x_1-x_2)^2}{x_1+x_2} 
+ (1-2\ep) \left(x_1+x_2-\f{s_{12}}{s_{123}}\right)
\right] \;\;.
\end{equation}
\begin{align}
\label{idensf}
 \Ph^{({\rm id})}_{{\bar q}_1q_2q_3} \,
&= C_F \left( C_F-\f{1}{2} C_A \right)
 \Biggl\{ (1-\ep)\left( \f{2s_{23}}{s_{12}} - \ep \right)\nn\\
&+ \f{s_{123}}{s_{12}}\Biggl[\f{1+x_1^2}{1-x_2}-\f{2x_2}{1-x_3}
    -\ep\left(\f{(1-x_3)^2}{1-x_2}+1+x_1-\f{2x_2}{1-x_3}\right) 
- \ep^2(1-x_3)\Biggr] \nn\\
&- \f{s_{123}^2}{s_{12}s_{13}}\f{x_1}{2}\left[\f{1+x_1^2}{(1-x_2)(1-x_3)}-\ep
    \left(1+2\f{1-x_2}{1-x_3}\right)
    -\ep^2\right] \Biggr\} + (2\lra 3) \, ,
\end{align}
and the variable $t_{ij,k}$ is defined as
\begin{equation}
\label{tij}
t_{ij,k}\equiv2\f{x_is_{ik}-x_js_{ik}}{x_i+x_j}+\f{x_i-x_j}{x_i+x_j} s_{ij}\, .
\end{equation}
The splitting function for the $q\to qgg$ decay can be
decomposed according to the different colour coefficients:
\begin{equation}
\label{qggsf}
 \Ph_{g_1 g_2 q_3} \, =
C_F^2 \, \Ph_{g_1 g_2 q_3}^{({\rm ab})}  \,
+ \, C_F C_A \,\Ph_{g_1 g_2 q_3}^{({\rm nab})}  \;\;,
\end{equation}
and the abelian and non-abelian contributions are
\begin{align}
\label{qggabsf}
 \Ph_{g_1 g_2 q_3}^{({\rm ab})} \, 
&=\Biggl\{\f{s_{123}^2}{2s_{13}s_{23}}
x_3\left[\f{1+x_3^2}{x_1x_2}-\ep\f{x_1^2+x_2^2}{x_1x_2}-\ep(1+\ep)\right]\nn\\
&+\f{s_{123}}{s_{13}}\Biggl[\f{x_3(1-x_1)+(1-x_2)^3}{x_1x_2}+\ep^2(1+x_3)
-\ep (x_1^2+x_1x_2+x_2^2)\f{1-x_2}{x_1x_2}\Biggr]\nn\\
&+(1-\ep)\left[\ep-(1-\ep)\f{s_{23}}{s_{13}}\right]
\Biggr\}+(1\lra 2) \;\;,
\end{align}
\begin{align}
\label{qggnabsf}
 \Ph_{g_1 g_2 q_3}^{({\rm nab})}  \,
&=\Biggl\{(1-\ep)\left(\f{t_{12,3}^2}{4s_{12}^2}+\f{1}{4}
-\f{\ep}{2}\right)+\f{s_{123}^2}{2s_{12}s_{13}}
\Biggl[\f{(1-x_3)^2(1-\ep)+2x_3}{x_2}\nn\\
&+\f{x_2^2(1-\ep)+2(1-x_2)}{1-x_3}\Biggr]
-\f{s_{123}^2}{4s_{13}s_{23}}x_3\Biggl[\f{(1-x_3)^2(1-\ep)+2x_3}{x_1x_2}
+\ep(1-\ep)\Biggr]\nn\\
&+\f{s_{123}}{2s_{12}}\Biggl[(1-\ep)
\f{x_1(2-2x_1+x_1^2) - x_2(6 -6 x_2+ x_2^2)}{x_2(1-x_3)}
+2\ep\f{x_3(x_1-2x_2)-x_2}{x_2(1-x_3)}\Biggr]\nn\\
&+\f{s_{123}}{2s_{13}}\Biggl[(1-\ep)\f{(1-x_2)^3
+x_3^2-x_2}{x_2(1-x_3)}
-\ep\left(\f{2(1-x_2)(x_2-x_3)}{x_2(1-x_3)}-x_1 + x_2\right)\nn\\
&-\f{x_3(1-x_1)+(1-x_2)^3}{x_1x_2}
+\ep(1-x_2)\left(\f{x_1^2+x_2^2}{x_1x_2}-\ep\right)\Biggr]\Biggr\}
+(1\lra 2) \;\;.
\end{align}
When a gluon decays collinearly, spin-correlations are present.
Here we are concerned only with spin-averaged splitting functions.
When the gluon decays in a $q\qb$ pair plus a gluon the splitting function is
\begin{equation}
\label{gqqsf}
\Ph_{g_1 q_2 {\bar q}_3}  \, =
C_F T_R \, \Ph_{g_1 q_2 {\bar q}_3}^{({\rm ab})} \,
+ \, C_A T_R\, \Ph_{g_1 q_2 {\bar q}_3}^{({\rm nab})}  \;\;,
\end{equation}
where
\begin{align}
\label{gqqabsfav}
 \Ph^{({\rm ab})}_{g_1q_2{\bar q}_3} \,&=
-2-(1-\ep)s_{23}\left(\f{1}{s_{12}}+\f{1}{s_{13}}\right)
+ 2\f{s_{123}^2}{s_{12}s_{13}}\left(1+x_1^2-\f{x_1+2x_2 x_3}{1-\ep}\right) 
\nn\\
&-\f{s_{123}}{s_{12}}\left(1+2x_1+\ep-2\f{x_1+x_2}{1-\ep}\right)
- \f{s_{123}}{s_{13}}\left(1+2x_1+\ep-2\f{x_1+x_3}{1-\ep}\right)
\;,
\end{align}
and
\begin{align}
\label{gqqnabsfav}
 \Ph^{({\rm nab})}_{g_1q_2{\bar q}_3}
\,&=\Biggl\{-\f{t^2_{23,1}}{4s_{23}^2}
+\f{s_{123}^2}{2s_{13}s_{23}} x_3
\Biggl[\f{(1-x_1)^3-x_1^3}{x_1(1-x_1)}
-\f{2x_3\left(1-x_3 -2x_1x_2\right)}{(1-\ep)x_1(1-x_1)}\Biggr]\nn\\
&+\f{s_{123}}{2s_{13}}(1-x_2)\Biggl[1
+\f{1}{x_1(1-x_1)}-\f{2x_2(1-x_2)}{(1-\ep)x_1(1-x_1)}\Biggr]\nn\\
&+\f{s_{123}}{2s_{23}}\Biggl[\f{1+x_1^3}{x_1(1-x_1)}
+\f{x_1(x_3-x_2)^2-2x_2x_3(1+x_1)}
{(1-\ep)x_1(1-x_1)}\Biggr] \nn\\
&-\f{1}{4}+\f{\ep}{2}
-\f{s_{123}^2}{2s_{12}s_{13}}\Biggl(1+x_1^2-\f{x_1+2x_2x_3}{1-\ep}
\Biggr) \Biggr\}
+ (2\lra  3) \, .
\end{align}
In the case of a gluon decaying into three collinear gluons we have:
\begin{align}
\label{gggsfav}
 \Ph_{g_1g_2g_3} \,&= C_A^2\Biggl\{\f{(1-\ep)}{4s_{12}^2}
t_{12,3}^2+\f{3}{4}(1-\ep)+\f{s_{123}}{s_{12}}\Biggl[4\f{x_1x_2-1}{1-x_3}
+\f{x_1x_2-2}{x_3}+\f{3}{2} +\f{5}{2}x_3\nn\\
&+\f{\left(1-x_3(1-x_3)\right)^2}{x_3x_1(1-x_1)}\Biggr]
+\f{s_{123}^2}{s_{12}s_{13}}\Biggl[\f{x_1x_2(1-x_2)(1-2x_3)}{x_3(1-x_3)}
+x_2x_3 -2 +\f{x_1(1+2x_1)}{2}\nn\\
&+\f{1+2x_1(1+x_1)}{2(1-x_2)(1-x_3)}
+\f{1-2x_1(1-x_1)}{2x_2x_3}\Biggr]\Biggr\}
+ (5\mbox{ permutations}) \;\;.
\end{align}


\end{document}